\documentclass[12pt,preprint]{aastex3}
\usepackage{epsfig}
\usepackage{psfig}

\usepackage[encapsulated]{CJK}
\usepackage[utf8]{inputenc}

\newcommand{\vvec}{{\mathbf v}}

\newcommand{\bvec}{{\mathbf B}}
\newcommand{\evec}{{\mathbf E}}

\newcommand{\be}{\begin{equation}}
\newcommand{\ee}{\end{equation}}
\newcommand{\bea}{\begin{eqnarray}}
\newcommand{\eea}{\end{eqnarray}}
\newcommand{\beax}{\begin{eqnarray*}}
\newcommand{\eeax}{\end{eqnarray*}}
\newcommand{\ba}{\begin{array}}
\newcommand{\ea}{\end{array}}
\newcommand{\bed}{\begin{description}}
\newcommand{\ed}{\end{description}}
\newcommand{\blc}{\begin{list}{$\circ$}{}}
\newcommand{\blb}{\begin{list}{$\bullet$}{}}
\newcommand{\el}{\end{list}}
\newcommand{\ben}{\begin{enumerate}}
\newcommand{\een}{\end{enumerate}}

\def\gsim{\,\lower3pt\hbox{$\sim$}\llap{\raise2pt\hbox{$>$}}\,}
\def\lsim{\,\lower3pt\hbox{$\sim$}\llap{\raise2pt\hbox{$<$}}\,}

\newcommand{\hatn}{\,\hat{\mathbf n}}
\newcommand{\hatx}{\,\hat{\mathbf x}}
\newcommand{\hatl}{\,\hat{\mathbf \ell}}
\newcommand{\hatb}{\,\hat{\mathbf b}}
\newcommand {\xvec}{{\mathbf x}}
\newcommand {\Svec}{{\mathbf S}}

\begin{document}

\title{A Magnetic Calibration of Photospheric Doppler Velocities}

\author{Brian T. Welsch}
\affil{Space Sciences Laboratory, University of California, 
Berkeley, CA 94720-7450}

\author{George H. Fisher}
\affil{Space Sciences Laboratory, University of California, 
Berkeley, CA 94720-7450}

\author{
\begin{CJK}{UTF8}{gbsn}
Xudong Sun (孙旭东) 
\end{CJK}
}
\affil{W. W. Hansen Experimental Physics Laboratory, Stanford
  University, Stanford, CA 94305, USA}


\begin{abstract}
The zero point of measured photospheric Doppler shifts is
uncertain for at least two reasons: instrumental variations (from,
e.g., thermal drifts); and the convective blueshift, a known
correlation between intensity and upflows.
Accurate knowledge of the zero point is, however, 
useful for (i) improving estimates of the Poynting flux of magnetic
energy across the photosphere, and (ii) constraining processes
underlying flux cancellation, the mutual apparent loss of magnetic
flux in closely spaced, opposite-polarity magnetogram features.
We present a method to absolutely calibrate line-of-sight (LOS)
velocities in solar active regions (ARs) near disk center using three
successive vector magnetograms and one Dopplergram coincident with the
central magnetogram.  It exploits the fact that Doppler shifts
measured along polarity inversion lines (PILs) of the LOS magnetic
field determine one component of the velocity perpendicular to the
magnetic field, and optimizes consistency between changes in LOS flux
near PILs and the transport of transverse magnetic flux by LOS
velocities, assuming ideal electric fields govern the magnetic
evolution.
Previous calibrations fitted the center-to-limb variation of Doppler
velocities, but this approach cannot, by itself, account for residual convective
shifts at the limb.
We apply our method to vector magnetograms of AR 11158, observed by
the Helioseismic and Magnetic Imager aboard the Solar Dynamics
Observatory, and find clear evidence of offsets in the Doppler zero
point, in the range of 50 -- 550 m s$^{-1}$.
In addition, we note that a simpler calibration can be determined from
an LOS magnetogram and Dopplergram pair from the median Doppler
velocity among all near-disk-center PIL pixels.
We briefly discuss shortcomings in our initial implementation,
and suggest ways to address these.  
In addition, as a step in our data reduction, we discuss use of temporal
continuity in the transverse magnetic field direction to correct
apparently spurious fluctuations in resolution of the 180$^\circ$
ambiguity.
\end{abstract}
%
%

\section{Introduction}
\label{sec:intro}

Solar variability is intimately related to magnetic flux at the solar
photosphere: solar flares, coronal mass ejections (CMEs), and enhanced
radiation from solar activity, ranging from radio to X-ray
wavelengths, all occur in the outer solar atmosphere above magnetized
regions of the photosphere.  Fundamentally, these manifestations of
solar activity are driven by the release of energy stored in or
transmitted by magnetic fields.

\subsection{Doppler Shifts \& the Poynting Flux}
\label{subsec:poynting}

Magnetic energy passes from the solar interior into the Sun's outer
atmosphere as an outward-directed Poynting flux $\Svec$ of magnetic
energy across the photosphere,
\be S_n = \hatn \cdot \Svec = c \hatn \cdot (\evec \times \bvec)/4 \pi 
\label{eqn:poynting0} ~, \ee 
where $c$ is the speed of light, $\evec$ and $\bvec$ are the
photospheric electric and magnetic fields, and $\hatn$ is the
outward-directed unit vector normal to the photosphere.  We will refer
to vectors perpendicular to the local normal as horizontal.  We use
the photosphere as the boundary between the solar interior and outer
atmosphere primarily for convenience, because the vector magnetic
field is only routinely measured at the photosphere.  A different
atmospheric layer could be used if observations of the vector magnetic
field were routinely produced there (see, e.g.,
\citealt{Metcalf1995}).

While $\bvec$ at the photosphere can be measured by vector
magnetographs, $\evec$ must be inferred by other means.  A popular
approach has been to use observed magnetic evolution in sequences of
magnetograms to estimate the electric field by inverting the
finite-difference approximation to Faraday's law,
\be \frac{\Delta \bvec}{\Delta t} = -c (\nabla \times \evec)
\label{eqn:faraday} ~, \ee 
where $\Delta t$ is the difference between the times $t_f$ and $t_i$
of final and initial magnetograms, respectively, with $\evec$
representing an average electric field over $\Delta t$.  Note that for
any $\evec$ which satisfies this expression, $(\evec - \nabla \psi)$,
where $\psi$ is an arbitrary scalar potential function, will also
satisfy it.  Hence, Faraday's law does not fully constrain $\evec$.

We assume the magnetic field is typically frozen to the plasma at the
photosphere on length scales observable with HMI, which has a pixel
size $\Delta x \simeq 0.5 \arcsec$, or $\simeq 360$ km near disk
center.  \citet{Kubat1986} estimated the magnetic diffusivity
$\eta_{\rm Ku}$ from collisions at the photosphere to be $10^8$ cm$^2$
s$^{-1}$ in magnetized regions with $B = 100$ G, implying the fluid
could slip across the magnetic field with relative velocity $u \sim
\eta_{\rm Ku}/\Delta x \simeq 3$ cm s$^{-1}$, which is completely
negligible compared to typical photospheric speeds.  Ignoring
diffusivity, the ideal Ohm's law relates photospheric velocities,
$\vvec$, to the electric field,
\be c \evec = -\vvec \times \bvec
\label{eqn:ideal} ~. \ee 
This implies estimates of $\vvec$ can be used to determine the flux of
magnetic energy (and magnetic helicity) across the photosphere (e.g.,
\citealt{Demoulin2003, Schuck2006}).  Several techniques have been
developed to estimate photospheric flows from $\Delta B_n/\Delta t$,
e.g., \citet{Chae2001, Kusano2002, Welsch2004, Schuck2006, Fisher2008} and
\citet{Schuck2008}.  These techniques are, however, imperfect
\citep{Rieutord2001, Welsch2007, Schuck2008}, so efforts to improve
them are ongoing.

\citet{Fisher2010} presented a method to determine $\evec$ from a
sequence of vector magnetograms using a poloidal-toroidal
decomposition (PTD) of the magnetic field, with Faraday's law.
Notably, the PTD method uses evolution of both the normal magnetic
field, $\Delta B_n/\Delta t$, and the normal electric current, $\Delta
J_n/\Delta t$, to estimate $\evec$.  While the PTD approach does not
rely upon the ideal Ohm's law to estimate $\evec$, any Ohm's law with
known resistive terms, including the ideal case, can be imposed {\em
  post facto} to constrain the solution for $\evec$.

For the ideal case, \citet{Fisher2012b} recently presented a method to
use Doppler measurements of the line-of-sight (LOS) component of the
velocity, $v_{LOS}$, to better constrain $\evec$.  They tested this
approach using synthetic magnetograms and Doppler data from an MHD
simulation.  They noted that flows along $\bvec$ (commonly referred to
as parallel or siphon flows) do not contribute to the ideal electric
field in equation (\ref{eqn:ideal}), but can contribute to $v_{LOS}$
in regions where the LOS component of the magnetic field is nonzero
($B_{LOS} \ne 0$).  Hence, they only incorporated Doppler data from
areas: (i) near polarity inversion lines (PILs), loci where the
component of the magnetic field along their (assumed) LOS changes sign
and the LOS field vanishes; and (ii) where the field transverse to the
LOS is large.  Their tests demonstrated that including Doppler data
near PILs substantially improves estimation of both $\evec$ and the
normal Poynting flux $S_n$.  For ideal evolution, this makes sense
because, in principle, the Doppler electric field $\evec^D$ from the
LOS velocity $v_{\rm LOS}$ and transverse magnetic field $\bvec_{\rm
  trs}$ along a PIL,
\be c \evec^D \equiv -(v_{\rm LOS} \times \bvec_{\rm trs} ) \vert_{\rm PIL} ~, \ee
is not uncertain by the gradient of a scalar potential, as are
estimates of $\evec$ from equation (\ref{eqn:faraday}) alone.

In real magnetograms, procedures have been developed to automatically
identify PILs (also sometimes called neutral lines; e.g.,
\citealt{Falconer2003}) of both the LOS components (e.g.,
\citealt{Schrijver2007, Welsch2008b, Welsch2009}) and normal
components \citep{Falconer2003} of magnetogram fields.

One factor hampering studies of Poynting fluxes has been the relative
dearth of sequences of vector magnetograms.  The launch of the
SpectroPolarimeter (SP) instrument with the Solar Optical Telescope
(SOT; \citealt{Tsuneta2008}) aboard the {\em Hinode} satellite
\citep{Kosugi2007} provided some seeing-free magnetogram sequences for
investigations of PIL dynamics, such as observations suggestive of an
emerging flux rope reported by \citet{Okamoto2008}.  However, the
cadence of the SP instrument is relatively slow compared to timescales
of photospheric evolution on arcsecond scales, and the field of view
(FOV) of SOT is limited. Vector magnetograms of active region fields
at higher cadence and over larger FOVs should be routinely produced by
the Helioseismic and Magnetic Imager (HMI) instrument
\citep{Scherrer2012} aboard the Solar Dynamics Observatory (SDO) and
the SOLIS vector spectromagnetograph (VSM; \citealt{Keller2003}),
enabling routine estimates of Poynting flux.

\subsection{Emergence \& Cancellation Along PILs}
\label{subsec:cancel}

Given the association between solar activity and magnetic fields, a
quantitative description of processes responsible for the introduction
and removal of magnetic flux into the solar atmosphere is essential to
understand solar activity.

\begin{figure}[t] 
  \centerline{\psfig{figure=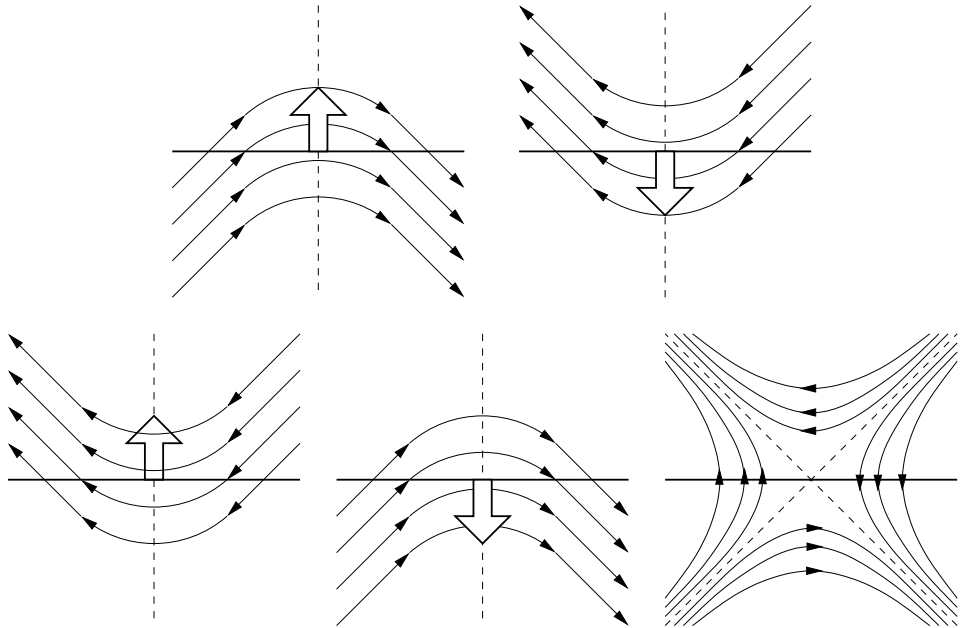,width=5.0in,clip=true}}
\caption{\footnotesize \textsl{Top row: sketch of field lines (thin,
    solid) in two possible processes that cause more flux to appear in
    magnetograms, $\Omega$-loop emergence (left) and U-loop
    submergence (right), viewed in a plane perpendicular to the
    polarity inversion line (PIL; vertical, dashed line in all
    panels).  The horizontal solid line represents the magnetogram
    layer.  Near disk center, if the evolution is ideal, then the
    plasma velocity (fat, white arrows) should produce Doppler shifts
    along the PIL.  Bottom row: Sketches of three possible processes
    in magnetic flux cancellation: emergence of U-loops (left),
    submergence of $\Omega$-loops (middle), or reconnective
    cancellation (right).  As with the top row, if the evolution is
    ideal, then the plasma velocity (fat, white arrows) should produce
    Doppler shifts along the PIL for emergence or submergence flux
    (left and middle panels, respectively) near disk center.  In
    principle, processes that increase or decrease total unsigned flux
    only occur along PILs.}}
\label{fig:all5}
\end{figure}

New magnetic flux typically appears at the solar photosphere via
emergence of $\Omega$-shaped loops from the convection zone (Figure
\ref{fig:all5}, top row, left panel).  About 3000 active regions are
typically cataloged by NOAA (which requires each to contain a visible
sunspot umbra) over one 11-year cycle of solar activity, each with
$\sim 10^{22}$ Mx ($10^{14}$ Wb) in unsigned flux.  Estimates vary for
the time required for emergence to replace the small-scale magnetic
fields in the quiet-Sun network, with unsigned flux on the order of
$\sim 10^{23}$ Mx over the photospheric surface, but they are
generally on the order of a day or less (e.g., Hagenaar et al. 2003).
\nocite{Hagenaar2003} We note that new flux can also appear in
photospheric magnetograms via the submergence of the bases of U-shaped
loops from above, as seen in simulations by \citet{Abbett2007} (Figure
\ref{fig:all5}, top row, right panel), though this process is not
often discussed.  Hence, the term {\em appearance} encompasses both
emergence and submergence.  This definition of the term differs
substantially from that used by \citet{Lamb2008}, in which appearance
refers to initial identification of a {\em feature} by a tracking
algorithm, where the definition of ``feature'' is algorithm-dependent.

Much like the appearance of new flux, the removal of magnetic flux
from the photosphere also plays a central role in solar activity.
Clearly, for the large-scale solar dynamo to operate cyclically, flux
that is introduced to the photosphere must eventually be removed or
recycled.  Similarly, flux emergence in the small-scale dynamo must be
statistically balanced by flux removal \citep{Schrijver1997}.

How is flux removed from the photosphere?  The short answer is ``flux
cancellation,'' which \citet{Livi1985} defined in observational terms
as ``the mutual apparent loss of magnetic flux in closely spaced
features of opposite polarity'' in magnetogram sequences.  Physically,
cancellation could correspond to (i) the emergence of U-shaped
magnetic loops (e.g., \citealt{Lites1995, vanDriel-Gesztelyi2000}),
(ii) the submergence of $\Omega$-shaped loops (e.g.,
\citealt{Rabin1984b, Harvey1999, Chae2004, Iida2010}), or (iii)
reconnection in the magnetogram layer (e.g., \citealt{Yurchyshyn2001,
  Kubo2007}; see also \citealt{Welsch2006}).  These possibilities are
sketched in the bottom row of Figure \ref{fig:all5}.

\citet{Spruit1987} proposed that opposite polarities within active
regions could be connected by U-loop subsurface extensions, and that
flux in these extensions might emerge as weak, ``sea-serpent''
(undulating) fields between strong field regions, eventually canceling
with active region flux via U-loop emergence.  \citet{Low2001}
proposed a process that also removes most active region flux from the
photosphere by the emergence of U-loops.  He suggests that the flux
rope that forms an active region contains many turns, such that a
magnetogram can intersect the flux rope many times after the flux rope
has partially emerged, and as the flux rope continues to emerge, much
of the flux cancels by U-loop emergence, with successive magnetograms
showing less and less flux remaining until little is left.  In
contrast to these models, \citet{vanBallegooijen2007} and
\citet{vanBallegooijen2008} constructed numerical models of active
region flux systems remaining anchored near the base of the convection
zone, with most canceling flux retracting to ``repair'' the toroidal
magnetic field deep in the solar interior.

\citet{Spruit1987} noted that sea-serpent cancellation might occur on
small scales. On unobservably small scales, this would produce
apparent {\it in situ} flux disappearance \citep{Wallenhorst1982a,
  Wallenhorst1982b, Gaizauskas1983}.  Further, unlike the
``self-cancellation'' of active region flux modeled by
\citet{Spruit1987}, \citet{Low2001}, and \citet{vanBallegooijen2007},
active region fields might also cancel with the ubiquitous,
small-scale fields of the quiet sun (e.g., Lin \& Rimmele 1999, Harvey
et al. 2007), \nocite{Lin1999, Harvey2007} including unresolved fields
\citep{SanchezAlmeida2009}.  This process would formally remove equal
amounts of active-region and quiet-Sun flux from the photosphere, but
if the latter were below a given magnetograph's noise threshold, it
would be undetectable, and the active-region flux would have seemed to
disappear.  While case studies of cancellation have been made, it is
unclear how much active region flux cancels on observable scales over
a solar cycle.  Hence, it is possible that cancellation with quiet-Sun
flux is the dominant method of active region flux removal.  Moving
magnetic features in moat flows around sunspots can explain rates of
sunspot flux loss (e.g., \citealt{Kubo2008}), but much flux in active
regions lies outside sunspots, and its disappearance must still be
understood.


Because magnetic fields are divergence-free, all field lines (tangent
lines of the vector field $\bvec$) form closed loops.  (These might be
infinitely long, or ergodic, but in any case field lines do not end.)
Magnetic flux can therefore only emerge or cancel where magnetic
fields are tangent to the photosphere.  These loci correspond to
normal-field PILs, where the normal magnetic field vanishes and
regions of positive and negative flux, where emerging (or submerging)
field lines thread the photosphere, are nearby.  In our terminology,
the appearance (or cancellation) of magnetic flux increases (resp.,
decreases) the total unsigned magnetic flux at the photosphere.
%

If the photospheric electric field during cancellation is ideal or
nearly so, then measurements of time-averaged Doppler shifts along
PILs should be able to distinguish between cancellation via either
U-loop emergence or $\Omega$-loop submergence.  Lack of a clear
Doppler signal while LOS flux cancels would be consistent with
reconnective cancellation.  We note that the magnetic field along PILs
away from disk center in LOS magnetograms can have a component that is
normal to the photosphere, implying LOS PILs away from disk center do
not exactly correspond to sites of flux appearance (or cancellation).
Hence, only Doppler shifts along LOS PILs near disk center can
effectively constrain the physical processes at work in cancellation.

Several case studies of Doppler shifts at cancellation sites have been
undertaken.  \citet{Yurchyshyn2001} and \citet{BellotRubio2005}
reported Doppler shifts consistent with upflows at the cancellation
sites they studied, which they interpreted as outflows from
reconnective cancellation.
%
%
\citet{Chae2004} and \citet{Iida2010} found evidence for flux
submergence during cancellation.  \citet{Kubo2007} studied
cancellations along several PILs in ASP magnetograms, and reported
mixed Doppler signals, which they interpreted in terms of reconnective
cancellation at multiple heights.  To determine the rest wavelengths
used to compute Doppler velocities, \citet{Chae2004} used quiet-Sun
values of line center, while \citet{Kubo2007} and \citet{Iida2010}
estimated their rest wavelengths by averaging line centers over their
FOVs.  As described below, however, these approaches to
determining rest wavelengths probably yield biased estimates.

\subsection{Biases in Doppler Shifts Along PILs}
\label{subsec:convective}


One factor complicating use of Doppler data both to estimate
photospheric electric fields and to constrain dynamics in flux
cancellation is inaccuracies in determination of the rest wavelengths
of photospheric lines in active regions.  Uncertainties in rest
wavelength can arise from both instrumental effects (e.g., from thermal
variations in components) and biases in analysis techniques.
In addition, the gravitational redshift is present \citep{Takeda2012}.
Periodicities in magnetic fields estimated by HMI \cite{Liu2012b}
on orbital time scales (12 and 24 hr) suggest
instrumental effects are present in estimated magnetic fields.  It is
plausible that similar effects should be present in
Doppler signals, a point we revisit below.

Because the position of line center is typically computed from
sampling predominantly quiet-Sun regions, where line profiles are
systematically shifted blue-ward by the convective blueshift
\citep{Dravins1981, Cavallini1986, Hathaway1992, Asplund2003,
  Schuck2010}, Doppler shifts in active regions typically exhibit
``pseudo-redshifts.''  The blue-ward bias of quiet-Sun line profiles
occurs because rising plasma is both (i) brighter than sinking plasma
(since rising plasma is hotter) and (ii) occupies a greater fraction
of an instrument's FOV than sinking plasma (since upwelling
convective cells are larger than downflow lanes).  Consequently, a
determination of line center position based upon the statistical
properties (e.g., means or medians) of Doppler images (Dopplergrams)
is biased by the upward motion of quiet-Sun plasma.  Observations and
modeling, by, e.g., \citet{Gray2009} and \citet{Asplund2003},
respectively, suggest that the magnitude of this bias can range from a
few hundred m sec$^{-1}$ to nearly 1 km sec$^{-1}$ for various
photospheric lines.  Because active region magnetic fields inhibit
convection (e.g., \citealt{Welsch2012}), measured line centers in
active regions are red-shifted relative to any rest wavelength derived
from quiet-Sun Doppler measurements. (Helioseismology requires
accurate measurement of {\em changes} in Doppler shifts, so is
insensitive to errors in determination of the rest wavelength.)
P. Scherrer (private communication 2009) has stated that one must
account for the convective blueshift to accurately determine Doppler
shifts in active regions.

\begin{figure}[!htb] 
  \centerline{\psfig{figure=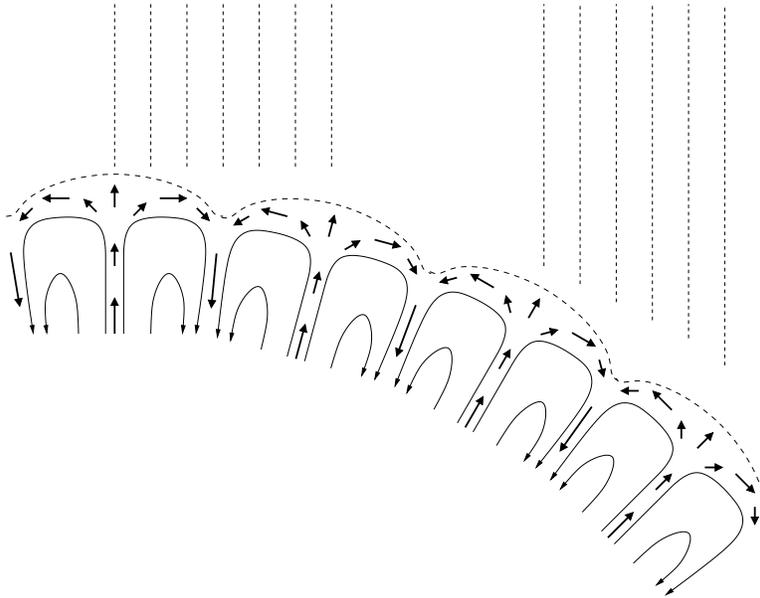,width=4.0in}}
\caption{\footnotesize \textsl{Idealized convective motions near the
    photosphere from center to limb.  Thick black arrows represent
    velocity vectors.  Thin black solid lines represent stream
    lines. Thick dashed line represents optical depth unity, and shows
    granules' three-dimensional structure (e.g., \citealt{Berger2007,
      Carlsson2004}).  For an observer looking from the top of the
    figure down, Doppler velocities along lines of sight (thin
    vertical dashed lines) both at disk center and toward the limb
    intersect more areas with velocities toward the observer than
    away.  This is one reason that determination of the rest
    wavelength by averaging (or taking the median) of line center
    positions over the FOV will result in an estimate biased
    blue-ward of the true rest wavelength.  This analysis ignores the
    additional factor that upflowing plasma is brighter than
    downflowing plasma, so the former contributes more to
    intensity-averaged line profiles.}}
\label{fig:c2l}
\end{figure}

By fitting measured Doppler shifts over the disk with profiles that
account for differential rotation, meridional flow, and center-to-limb
variations, any overall constant Doppler shift (sometimes referred to
as the convective limb shift; \citealt{Hathaway1992}) can be estimated
\citep{Snodgrass1984, Hathaway1992, Schuck2010}.  This approach,
however, has a major physical uncertainty: the physics of the
center-to-limb variation in average Doppler shift in the particular
spectral line used by HMI (or any other spectral line) involves
detailed interactions between height of formation, the height of
convective turnover, the variation with viewing angle of the average
convective flow speed, and the variation with viewing angle of optical
depth \citep{Carlsson2004, Takeda2012}.  We note that diverging flows tangent to
the photosphere in granules can, depending upon optical depth in
granules at the formation height of the line, produce a convective
blueshift toward the limb, because diverging flows on the near sides
of granules approach the viewer, while receding flows on the far sides
of the granules are at least partially obscured by the optical depth
of the granules.  This is sketched in Figure \ref{fig:c2l}.  Hence,
fitting the observed center-to-limb variation in line-center positions
does not imply that all bias from convective motions has been
determined.  In principle, simulations of convection near the
photosphere (e.g., \citealt{Asplund2003, Fleck2011}) could be used to
characterize the expected center-to-limb variation in line centers,
which could then be used to remove the modeled convective blueshift
bias (but not any instrumental biases).  We are unaware of published
comparisons of observed and modeled center-to-limb variations in the
HMI spectral line.

The goal of this paper is to describe a ``magnetic
calibration'' technique to estimate any spatially uniform bias in
measured Doppler velocities. In basic terms, this method estimates
such a bias by requiring statistical consistency between two
independent measures of changes in flux near LOS-field PILs, the loci
where flux emerges and submerges: (1) $\Delta \Phi_{\rm LOS}$, half
the change of total unsigned LOS flux near PILs over a time interval
$\Delta t$; and (2) $\Delta \Phi_{vBL}$, the amount of transverse flux
transported upward or downward across the magnetogram layer over
$\Delta t$, computed by summing the Doppler electric field --- the
product of the Doppler velocity and transverse magnetic field strength
--- along PILs.  This consistency relies upon both Faraday's law and
the assumption of ideal evolution.

The rest of the paper is organized as follows.  In the next section,
we describe our calibration method in greater detail.  We then apply
the method to the initial vector magnetogram sequence released by the
HMI Team, starting with a description of our preliminary treatment of
the data (\S \ref{sec:data}) followed by a step-by-step demonstration
of the method (\S \ref{sec:demo}) and analysis of its results (\S
\ref{sec:results}).  We conclude with a brief summary in \S
\ref{sec:summ}.

\section{Calibrating Doppler Shifts with Faraday's Law} 
\label{sec:theory}

Consider a region of the photosphere where flux is either increasing
or decreasing due to appearance or cancellation (by any of the
mechanisms discussed above), respectively.  In an area $A$ containing
flux of one polarity, Faraday's law and Stokes' theorem relate the
time rate of change of magnetic flux, $\partial \Phi/\partial t$,
through $A$ with the integral of the electric field $c \evec$
projected onto a closed curve $\cal C$ bounding $A$,
\bea  \frac{\partial \Phi_A}{\partial t} 
&=& \int dA \, \frac{\partial \bvec}{\partial t} \cdot \hatn 
\label{eqn:dphidt} \\
&=& - \int dA \, (\nabla \times c \evec) \cdot \hatn \\
&=& - \oint_{\cal C} c \evec \cdot d\xvec_{\parallel}
~, \label{eqn:stokes0} \eea
where $\hatn$ is the unit normal vector on $A$ and $\hatx_{\parallel}$
is the tangent vector to ${\cal C}$ oriented to circulate in a
right-hand sense (counterclockwise) with respect to $\hatn$.  This
must apply separately to the areas $A_\pm$ containing the flux of each
polarity.

We restrict ourselves to the case where the flux through areas $A_\pm$
changes only due to emergence or cancellation, so flux is only added to
or removed from $A_\pm$ along the PIL.  Hence, the only contribution
to the integral of $\evec$ along $\cal C$ occurs along the PIL,
\be  \frac{\partial \Phi_A}{\partial t} 
= - \int_{\rm PIL} c \evec \cdot d\xvec_{\parallel}
~. \label{eqn:pil_e} \ee
The time rate of change of flux is therefore equal to the voltage drop
along the PIL.  This equivalence is exact, without approximation.  In
real data, however, some PILs might not be observable at a given
instrument's spatial resolution, flux cannot be measured without
uncertainties, and flows on the peripheries of $A_{\pm}$ can disperse
flux until its density falls below a magnetograph's sensitivity.  In
addition, we note that real PILs are often intermittent, and possess
complicated spatial structure, motivating our application of this
approach with real data, below.

Whether the electric field contains a non-ideal component or not, the
rate of ideal transport of magnetic flux across the PIL is given by
replacing $c \evec$ with $-(\vvec \times \bvec)$ in equation
(\ref{eqn:pil_e}),
\be  \frac{\partial \Phi_{vBL}}{\partial t} 
= \int_{\rm PIL} (\vvec \times \bvec) \cdot d\xvec_{\parallel}
~, \label{eqn:pil_v0} \ee
where the subscript $vBL$ denotes that this expression represents the
transport of flux across the PIL by plasma flows.  We now define a unit
vector perpendicular to the PIL,  $\hatx_{\perp}$, with respect to 
$\hatn$ and $\hatx_{\parallel}$,
\be \hatx_{\perp} = \hatx_{\parallel} \times \hatn ~, \label{eqn:hatx} \ee
which points in the direction of the horizontal gradient of $B_n$
evaluated at the PIL (because the PIL is assumed to be a zero contour
of $B_n$), and $d\xvec_\parallel = \hatx_{\parallel} dL$. Then
\bea  \frac{\partial \Phi_{vBL}}{\partial t} 
&=& \int_{\rm PIL} dL \, (v_n B_\perp - v_\perp B_n) \\
&=& \int_{\rm PIL} dL \, v_n B_\perp 
~, \label{eqn:pil_v} \eea
where $\bvec_h$ the field tangent to the surface, $B_\perp$ is the
component of $\bvec_h$ along $\hatx_{\perp}$, and the final equality
holds because $B_n$ vanishes along the PIL.

\begin{figure}[!htb] 
  \centerline{\psfig{figure=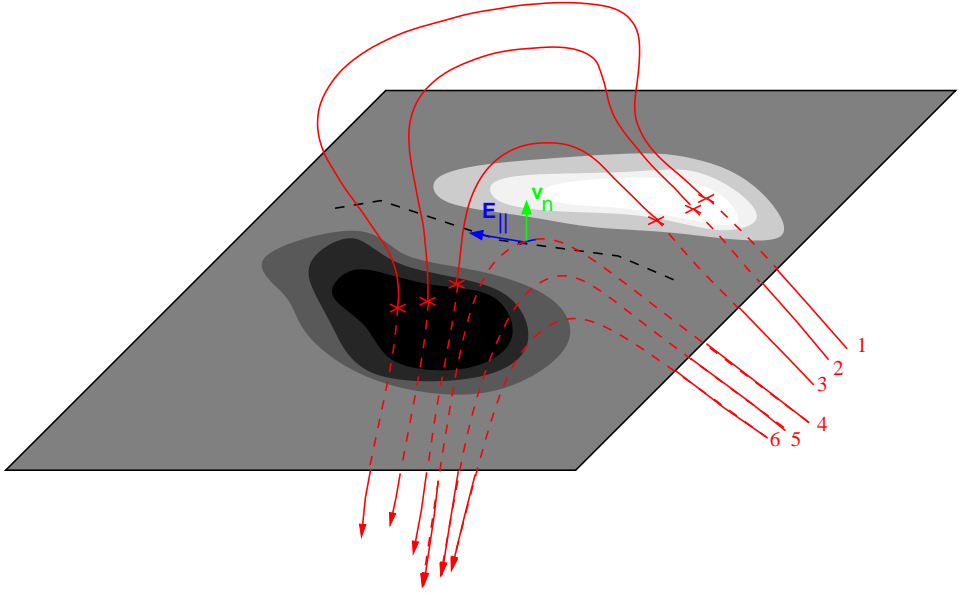,width=5.0in}}
\caption{\footnotesize \textsl{A depiction of ideal flux emergence
    across the photosphere (gray surface). Magnetic field lines 1,2,
    and 3 extend from the solar interior (red dashed lines), across
    the photosphere (red $\times$'s), and out into the outer solar
    atmosphere (red solid lines).  Along a PIL (dashed black line)
    between regions of positive (white/light shades) and negative
    (black/dark shades) magnetic flux, a velocity (green vector) normal
    to the photosphere is driving flux emergence.  At the instant
    depicted, the field line labeled 4 is crossing the
    photosphere. Additional vertical transport of flux (containing
    field lines 5 and 6) would cause unsigned flux through the
    positive and negative flux regions to increase.  An ideal electric
    field (blue vector) parallels the PIL at the emergence site.}}
\label{fig:pil_emrg}
\end{figure}

If the magnetic evolution is ideal, then the rate of change of flux in
each area $A_\pm$ will match the rate of ideal flux transport across
the PIL,
\bea \left \vert \frac{\partial \Phi_{A_\pm}}{\partial t} \right \vert &=& 
\left \vert \frac{\partial  \Phi_{vBL}}{\partial t} \right \vert \\
\left | \int dA_\pm \, \frac{\partial \bvec}{\partial t} \cdot \hatn \right |
&=& \left | \int_{\rm PIL} dL \, v_n B_\perp \right |
~, \label{eqn:ideal_equiv} \eea
where $dA_\pm$ denotes integration over either polarity (not both).
The key point here is that the PIL-integrated rate of transport of
horizontal flux by the normal velocity matches the rate of change of
normal flux through $A_\pm$.  Unlike equation (\ref{eqn:pil_e}), this
equivalence rests upon the assumption of ideality.  (Also, we discuss
the effect of filling factors below.)  In general, only the unsigned
changes in flux match, since both up- and downflows can lead to flux
decrease by cancellation (Figure \ref{fig:all5}, bottom left and
center), and both up- and downflows can lead to flux appearance, by
either $\Omega$-loop emergence or $U$-loop submergence (Figure
\ref{fig:all5}, top row; and \citealt{Abbett2007}).  Figure
\ref{fig:pil_emrg} sketches emergence of an $\Omega$-loop flux system,
one physical situation to which this formalism can be applied.

We can apply this formalism to PILs of the LOS field near disk center.
We require both Dopplergrams, in which the LOS velocity $v_{\rm LOS}$ is
measured, and vector magnetograms, in which pixel-averaged flux
densities of magnetic components both along the LOS, $B_{\rm LOS}$,
and transverse to the LOS, $\bvec_{\rm trs}$, are determined.  We
approximate the normal vector $\hatn$ used above with the unit vector
along the LOS, $\hatl$, replace the exact time derivatives above with
finite-difference approximations, and change integrations to
summations over pixels in $A_\pm$ and along the PIL. Equation
(\ref{eqn:dphidt}) then becomes
\be  \frac{\Delta \Phi_\pm}{\Delta t} 
= \sum_\pm \, (\Delta x)^2 \frac{B_{LOS}(t_f) - B_{LOS}(t_i)}{\Delta t} 
~, \label{eqn:dphidt_fd} \ee
where $\Delta x$ is the pixel length, and each sum --- there are two,
one over each polarity --- runs over pixels in the neighborhood of the
PIL.  (We address identification of PILs and definition of
neighborhoods near each PIL below.)
%
%
Equation (\ref{eqn:pil_v}) then becomes
\be  
\frac{\Delta \Phi_{vBL}}{\Delta t} 
= \sum_{\rm PIL}  \Delta x \, v_{\rm LOS} B_\perp 
\label{eqn:pil_v_fd} ~, \ee
where the sum runs over PIL pixels, and $B_\perp$ now refers to the
component of $\bvec_{\rm trs}$ along $\hatx_{\perp}$. Equation
(\ref{eqn:ideal_equiv}) then becomes, formally,
\bea | {\Delta \Phi_{LOS, \pm}} | &=& | {\Delta \Phi_{vBL}} |
\label{eqn:ideal_equiv_fd0} \\
\left | \sum_{+\,\,\mbox{or}\,\,-} \, (B_{\rm LOS}(t_f) - B_{\rm LOS}(t_i)) \right |
&=& \left | \Delta t \sum_{\rm PIL} \Delta x \, v_{\rm LOS} B_\perp \right |
\label{eqn:ideal_equiv_fd} ~, \eea
where we have multiplied equations (\ref{eqn:dphidt_fd}) and
(\ref{eqn:pil_v_fd}) by $\Delta t$ to deal with changes in flux
instead of rates of change in flux, and the sum in the left half of
equation (\ref{eqn:ideal_equiv_fd}) runs over the near-PIL
neighborhood of one LOS polarity or the other.

%
%

Crucially, LOS velocities along PILs are therefore constrained by
evolution of nearby LOS flux: over a time interval $\Delta t$, the
change in LOS flux in each polarity in the neighborhood of a PIL
should match, within methodological uncertainties, the transport of
transverse magnetic field by LOS velocities summed along PIL pixels.


As mentioned in the introduction, however, a nonzero bias velocity
$v_0$ might be present in estimated Doppler velocities from HMI,
perhaps due to instrumental effects or the convective blueshift. We
define the biased and true Doppler velocities as $v_{\rm LOS}'$ and
$v_{\rm LOS}$, respectively, which are related via
\be v_{\rm LOS}' = v_{\rm LOS} + v_0 ~. \ee 
We follow the astrophysical convention that receding velocities with
respect to the observer --- i.e., redshifts --- are positive. Then for
$v_0 > 0$, the biased Doppler shift is {\em more red} than the true
Doppler shift.  Such an offset in the Doppler velocity would mean that
the true rate of flux emergence or submergence is related to the
biased rate $\Delta \Phi'_{vBL}/\Delta t$ by
\bea  
\frac{\Delta \Phi_{vBL}}{\Delta t} 
&=& \sum_{\rm PIL}  \Delta x \, v_{\rm LOS}' B_\perp 
- \sum_{\rm PIL}  \Delta x \, v_0 B_\perp 
\label{eqn:phi_vbl_prime0}  \\
&=& \frac{\Delta \Phi_{vBL}'}{\Delta t} - \frac{\Delta \Phi_{\rm bias}}{\Delta t}
\label{eqn:phi_vbl_prime2}   
~, \eea
where the {\em bias flux} $\Delta \Phi_{\rm bias}$ is due to the bias
velocity $v_0$, 
\be \Delta \Phi_{\rm bias} \equiv v_0 \Delta t \overline{B L} 
~, \label{eqn:phi_bias} \ee
and where we define the {\em magnetic length} of the PIL as
\be \overline{B L} \equiv \sum_{\rm PIL} (\Delta x) B_\perp
~. \label{eqn:maglen} \ee 

Equation (\ref{eqn:ideal_equiv_fd0}) then implies
\be \left | \Delta \Phi_{\rm LOS, \pm} \right |
= \left | \Delta \Phi_{vBL}' - \Delta \Phi_{\rm bias} \right | 
~. \label{eqn:los_vbl_bias} \ee
In the absence of errors, the observed quantities $\Delta \Phi_{\rm
  LOS}$ and $\Delta \Phi_{vBL}'$ enable $\Delta \Phi_{\rm bias}$ to be
determined, from which the bias velocity $v_0$ can be found via
equation (\ref{eqn:phi_bias}).  Errors are certainly present in these
quantities, however, and the evolution also might not be ideal.  At
this point, however, two key points should be noted:
\blb
\item First, the rest wavelength of the spectral line used to infer
  Doppler velocities is {\em a unique, well-defined physical
    quantity.}  
\item Second, any bias in the rest wavelength should be {\em spatially
  uniform} over the instrument FOV in an individual
  measurement.
\el
These ideas imply that the {\em set} of bias fluxes $\{ \Delta
\Phi_{\rm bias} \}$ and magnetic lengths $\{ \overline{BL} \}$
measured over a set of PILs near disk center can be used to estimate
$v_0$ statistically, which also enables quantifying uncertainties in
the estimate.

It should be noted that, although we derived the flux-matching
constraint in equation (\ref{eqn:ideal_equiv_fd}) by considering flux
appearance / cancellation, {\em this constraint can determine $v_0$
  whether or not flux is emerging or submerging.}  Imagine, for
instance, no actual emergence / submergence (i.e., no significant
changes in total unsigned LOS flux) were occurring; a nonzero $v_0$
would, however, imply {\em spurious} emergence / submergence along
PILs.

As noted in the introduction, away from disk center, the approximate
coincidence between LOS PILs and the radial-field PILs where flux
appears or cancels breaks down.  While the formalism used here could
potentially be developed further to enable analysis of PILs
significantly away from disk center, we will restrict our analysis here to
PILs near disk center.  In Appendix \ref{app:offcm}, we consider
pathologies in applying this method to PILs away from disk center.


There are several ways one might estimate the flux changes $\Delta
\Phi_{\rm LOS}$ and $\Delta \Phi_{vBL}$ and magnetic length
$\overline{BL}$ for each PIL, along with several ways sets of these
estimates can be used to estimate $v_0$, and thereby remove any bias
present in measured Doppler velocities $v_{\rm LOS}'$.  Accordingly,
in the following sections, we demonstrate our approach with actual HMI
observations, with several goals in mind: first, to make the method
more clear; second, to show that the method can be used with real
data; third, to show the method derives physically reasonable values;
and fourth, to investigate the extent to which the method depends upon
input parameters and is susceptible to errors.

\section{HMI Observations}
\label{sec:data}

\begin{figure}[!htb] 
  \centerline{\psfig{figure=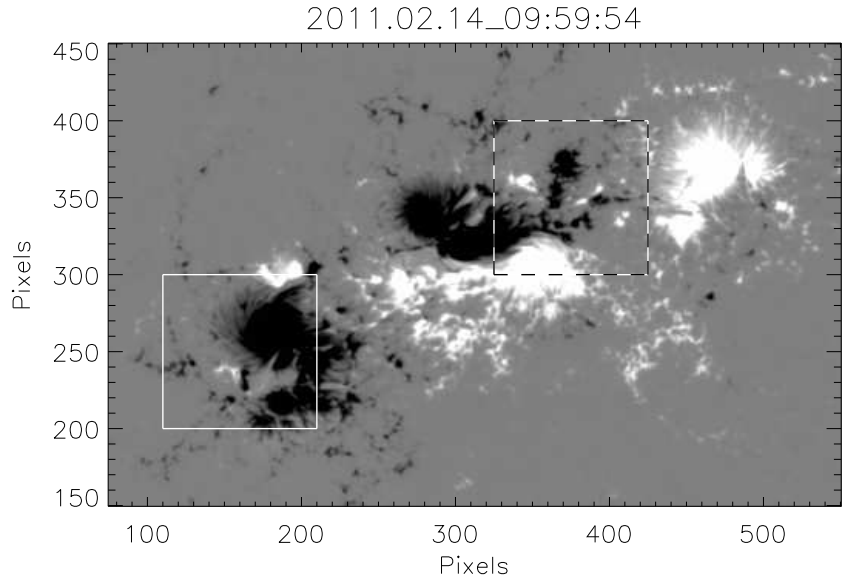,width=5.0in}}
\caption{\footnotesize \textsl{A grayscale image of the LOS magnetic
    field in AR 11158 near 10:00UT on 2011/02/14 in subregion of the
    data released by the HMI Team.  White/black represent
    positive/negative LOS flux, respectively.  Image saturation is set
    at $\pm 750$ G.  The solid white box outlines a subregion of the
    FOV shown in Figure \ref{fig:flips}, and the white/black dashed
    box outlines the subregion shown in Figure \ref{fig:pilvects}.}}
\label{fig:blos}
\end{figure}

Recently, the HMI Team released a sequence of ``cutout''
vector
magnetograms\footnote{ftp://pail.stanford.edu/pub/HMIvector/Cutout/}
and
Dopplergrams\footnote{ftp://pail.stanford.edu/pub/xudong/stage2/IcV/}
from NOAA AR 11158, from 12-16 Feb. 2011.  Members of the HMI Team
have put detailed information about this dataset
online,\footnote{http://jsoc.stanford.edu/jsocwiki/VectorPaper}, and
are preparing a paper describing production of this dataset.  This
active region was the source of an X2.2 flare on on 2011/02/15,
starting in GOES at 01:44, ending at 02:06, and peaking at 01:56.
Figure \ref{fig:blos} shows the LOS magnetic field in a subregion of
the data array near 10:00UT on 2011/02/14.

In this section, we briefly describe procedures that we
undertook to prepare this data prior to applying our calibration
procedure.\footnote{More detailed notes describing our procedures are
  online at \\
  http://solarmuri.ssl.berkeley.edu/$\sim$welsch/public/manuscripts/Doppler\_calib/hmi\_data\_notes\_current.txt}

After downloading the FITS data files, several processing steps were
required, which we describe here.  Because the \texttt{read\_sdo.pro}
IDL procedure did not (at the time of this writing) properly handle
pixels set to the BLANK value in the FITS headers, we used the
\texttt{fitsio\_read\_image.pro} procedure\footnote{Available online
  at http://www.mps.mpg.de/projects/seismo/GDC\_USE/using\_drms.html,
  along with a required, compiled shared-object file,
  \texttt{fitsio.so}.}  to read in the data.

\subsection{Cropping}
\label{subsec:crop}

To reduce the full dataset to a more manageable size, we focus on a
subset of the full five-day sequence. To study photospheric magnetic
evolution prior to the X flare, and to baseline this evolution against
post-flare evolution, we retain about 72 hours of data, from midnight
at the start of the 13th until midnight at the end of the 15th,
inclusive of endpoints.  Given the 12-minute cadence, the resulting
time series consists of 361 time steps.
The active region was at S19E11 at 00:30UT on 2011/02/13 and S21W27 at
00:30UT on 2011/02/16.
Pixels lacking data near the edges of the cutout FOV, from
artifacts of the cutout process, were cropped: columns [0 -- 24] and
rows [0 -- 5] were removed for all steps.  

\subsection{Removing SDO Motion and Solar Rotation}
\label{subsec:scrot}

Next, Doppler velocities were corrected for spacecraft motion.  Due to
SDO's geosynchronous orbit, its velocity along the radial,
Sun-observer line can be large, of order $\pm$ 5 km s$^{-1}$.  Also,
because the Earth is orbiting westward about the Sun, and SDO orbits
Earth, there is a significant projection of SDO's westward motion onto
lines of sight to many pixels.  In addition, there is a nonzero
component of SDO's northward velocity onto the LOS.  While the radial
component is larger than the W and N components, the latter can be
significant --- a few tens of m s$^{-1}$ or more. Accordingly, the
spacecraft velocity was projected onto the LOS to each pixel --- in
the small-angle approximation --- and subtracted off the measured
Doppler velocities.  Then Stonyhurst latitudes and longitudes of each
pixel (accounting for the solar $B$ and $P$ angles) were computed to
correct Doppler velocities for solar rotation, using the ``magnetic/ 2-day
lag'' rotation rate found by \citet{Snodgrass1983}, for which
\bea
\omega_{\rm rot} &=& A + B \sin(\tilde \theta)^2 + C \sin(\tilde \theta)^4 
\label{eqn:snod} \\
A &=&  2.902  \\
B &=& -0.464 \\
C &=& -0.328 
~, \eea
where $\omega_{\rm rot}$ is the photospheric rotation rate, $\tilde
\theta$ is latitude, and $A, B$, and $C$ are in microrad s$^{-1}$.
Accurate correction for rotation prior to application of our
calibration procedure is not necessary; but without removing it, the
bias velocity estimates will include (and could be dominated by) the
Doppler shift from rotation.

We also note that the Doppler signal from helioseismic $p$-mode
oscillations, which have a peak in power at frequencies near 1/300 Hz,
could introduce non-convective signals into the 720-second cadence
Dopplergrams.  While individual mode amplitudes are on the order of
a few tens of cm s$^{-1}$, the overall motion due to the superposed
modes can be 500 m s$^{-1}$ or more, which is comparable to convective
motions.  In practice, however, HMI's 720-second Dopplergrams are
created by averaging higher-cadence filtergrams with a
cosine-apodized, 1215-second boxcar \cite{Liu2012a} with a full width
at half maximum of 720 seconds (Schou 2012, private communication).
This effectively acts as a filter, which would reduce a random-phase
sinusoidal signal with period of 300 s and amplitude of 500 m s$^{-1}$
to an averaged value of less than 20 m s$^{-1}$ (less than the noise
level of HMI Dopplergrams).

\subsection{Removing Azimuth Reversals}
\label{subsec:filter}

\begin{figure}[!htb] 
  \centerline{\psfig{figure=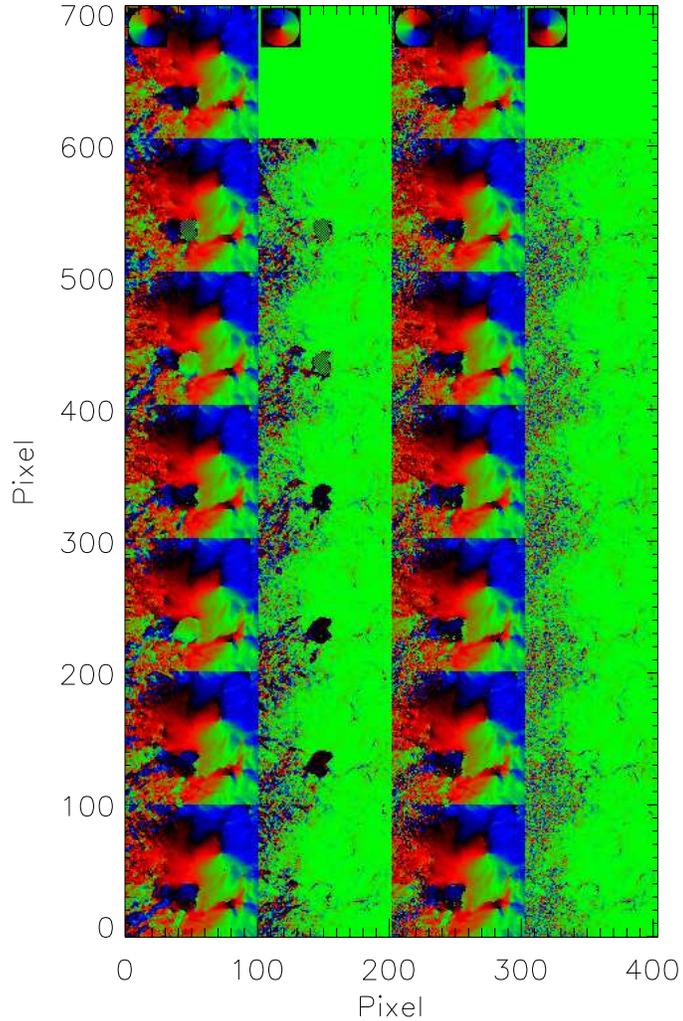,width=4.0in}}
\caption{\footnotesize \textsl{We applied a filtering procedure to
    remove back-and-forth flips in azimuth, which are likely due to
    frame-to-frame inconsistencies in ambiguity resolution.  From left
    to right, 1st column: 100 $\times$ 100 pixel maps of
    transverse-field azimuths (color wheel in top panel) from the
    solid white box in Figure \ref{fig:blos} from seven successive vector
    magnetograms (separated by 12 minutes), starting at 10:00UT on
    2011/02/14. Time increases down. 2nd column: Difference in azimuth
    between current and previous maps (note different color wheel in
    top panel).  Initial difference map is arbitrarily set to zero.
    3rd column: Filtered versions of the azimuth maps in the 1st
    column, in which continuity in azimuths has been imposed (color
    wheel in top panel). 4th column: Difference in azimuth between
    current and previous filtered maps (color wheel in top panel).
    Initial difference map is arbitrarily set to zero. Back-and-forth
    flips in many pixels have been removed.}}
\label{fig:flips}
\end{figure} 

Accurate estimation of electric fields by PTD (or, really, any method)
requires that changes in $\bvec$ between magnetograms arise from physical
processes on the Sun, as opposed to artifacts of the measurement
process.  Hence, any spurious changes in the measured field, where
they can be identified, should be removed.

We noticed that in successive images of the transverse-field azimuths,
the inferred azimuths in some regions flip by nearly 180 degrees from
one frame to the next, as would be expected from errors in resolving
the 180-ambiguity.  The seven successive frames (12 min. apart) in the
left column of Figure \ref{fig:flips} show an example of this effect,
with time increasing downward.  The area shown corresponds to that in
the solid white box in Figure \ref{fig:blos}, and the initial frame
corresponds to that in Figure \ref{fig:blos}.  Azimuths range over
[0,360] degrees.  The next column shows maps of the angular difference
(the interior angle) between the current and previous azimuths.
Differences in azimuths range from [-180,180] degrees (note the
rotated color wheel in the top panel).  The black patches in rows 4 --
6 of this column correspond to a region in which azimuths flip by
nearly 180$^\circ$ in a region from one frame to the next.  The
checkerboard pattern in the same region in rows 2 -- 3 of this column
implies a spatially alternating pattern of 180$^\circ$ azimuth flips,
and probably arises from lack of convergence in the simulated
annealing algorithm used to infer the azimuths \citep{Leka2009b}.
Also, the red-and-blue, speckled regions toward the left side of each
frame in this column exhibit rapid spatial variations in azimuth
changes.  These speckled regions correspond to areas with relatively
weak field strength ($B < 200$ G), where azimuth determinations could
be more problematic than in strong-field regions.

Interpreting frame-to-frame azimuth flips over finite regions as
spurious, we seek to identify and remove these artifacts.  Our goal is
to automate detection of such flips, and our basic approach is to
identify suspicious changes in azimuth, which we envision as ``top
hats'' (or inverted top hats) in the running differences of azimuths
in individual pixels: large, positive (or negative) jumps in azimuth
for one time step, followed by reversals to the pre-jump level at the
next step.  The actual patching is simpler: we add 180 degrees to the
azimuth, and output the flipped data modulo 360 degrees.

We tried more than one approach to detecting and correcting spurious
azimuth changes before finding one that we think works well.  To save
other researchers from repeating our efforts, we now describe, in
detail, both our failed approach and the approach we found worked
best.

In our initial, unsuccessful approach, we simply identified all
pixels with unsigned changes in azimuth from frame $(i-1)$ to frame
$i$ above a given threshold (we tried, e.g., 120, 135, 150, and 170
degrees), and flipping these in frame $i$.  The {\em updated} frame
$i$ was then used as a reference for finding large changes in frame
$(i+1)$.  Using this approach, once a pixel is flipped, its new state
tends to be propagated forward in time.  But legitimate magnetic
evolution can cause this state to be incorrect, leading to spatially
isolated pixels that differ from their neighbors.  This contradicts a
central tenet of the original ambiguity resolution procedure: currents
should be minimized.  Further, the number and spatial arrangement of
problematic pixels appears to depend strongly on threshold.

The best approach we found is to flip azimuths only in pixels where
both: (i) the jump in azimuth was larger than 120$^\circ$ between
frames; and (ii) the jump in azimuth increased the time-averaged,
unsigned changes in azimuth over a the set of frames within $\pm N$
steps of a given frame.  Use of a range of images on either side of
the $i$-th image is necessary to capture the essential character of a
top hat: a large change, followed by a reversal.  (A large azimuth
change from one frame to the next, by itself, might be legitimate ---
e.g., by horizontal advection of a region with a spatial transverse
field reversal.)  We determined suspicious flips by comparing means of
two arrays of $2N$ 
azimuth differences: (1) unsigned differences
between the azimuths at frame $i$ and azimuths in $\pm N$ neighboring
frames in time, $[i-N:i-1]$ and $[i+1:i+N]$; and (2) unsigned
differences between the {\em flipped} azimuths (hypothetical data,
with {\em all} pixels' azimuths flipped) at frame $i$ with actual
azimuths in $\pm N$ neighboring frames in time, $[i-N:i-1]$ and
$[i+1:i+N]$.  Pixels for which flipping would decrease the mean,
unsigned, frame-to-frame angular differences are then flipped.

We call frames in which azimuths have been flipped {\em filtered.}
Filtered azimuths are used for past frames, $[i-N:i-1],$ while
unfiltered azimuths are used for future frames $[i+1:i+N],$ so this
approach is not time-symmetric.  This approach should be able to deal
with top hats that are $N$ steps wide, but not wider. We tried both
$N=4$ and $N=2$, and show results with $N$ = 2 here. Using $N=4$
results in 2--5\% fewer flips in a given frame, so is slightly more
restrictive.  We did not, however, see much difference between results
from $N=2$ and $N=4$ in strong-field regions, so most differences are
probably in weak-field regions where inference of the direction of
transverse fields is less reliable anyway.

The third column from left in Figure \ref{fig:flips} shows filtered
data from the same seven successive frames as in the left-most column,
with suspicious changes in azimuths identified by our approach flipped
by 180$^\circ$.  The right-most column shows maps of the angular
difference between the current and previous azimuths (note the rotated
color wheel in the top panel) from the third column.  The checkerboard
and black patches visible in the second column are not present in this
column.  Also, changes in azimuths in the red-and-blue, speckled
regions on the left sides of each frame are not as large in this
column as in the second column.  This suggests our approach 
decreases fluctuations of azimuths in weak-field regions, too.


\section{Demonstration of Electromagnetic Calibration} 
\label{sec:demo}

We now seek to quantify any bias velocity in the Doppler velocity
measurements.  Since this will require associating pixel values in
successive frames (e.g., in equation \ref{eqn:dphidt_fd}), we first
co-aligned the plane-of-sky (POS) Dopplergram, field strength, and
inclination arrays.  We used the $B_{\rm LOS}$ arrays as the reference
observations, since structures in $B_{\rm LOS}$ are long-lived (e.g.,
\citealt{Welsch2012}).  Whole-frame shifts between each $B_{\rm LOS}$
array were determined to sub-pixel accuracy using Fourier
cross-correlation, and the data arrays were shifted accordingly via
Fourier interpolation.\footnote{See
  http://solarmuri.ssl.berkeley.edu/$\sim$welsch/public/software/shift\_data.pro}
Instead of wrapping data across image edges, due to the assumed
periodicity in the Fourier method, data shifted out of the image field
of view were zeroed out.  Interpolated field inclinations outside
[0,180] were capped at these values.  We then used the co-aligned
inclinations and field strengths to derive co-aligned LOS and
transverse fields.

Our calibration method requires several additional tasks: identifying
PILs of the LOS field; quantifying changes in LOS
magnetic flux near those PILs; summing transverse fields $B_{\rm trs}$
and Doppler velocities $v_{\rm LOS}'$ along identified PILs; and
statistically estimating any offset $v_0$ in these Doppler velocities.

\subsection{Identification of PILs}
\label{subsec:pils}

\begin{figure}[!htb] 
  \centerline{\psfig{figure=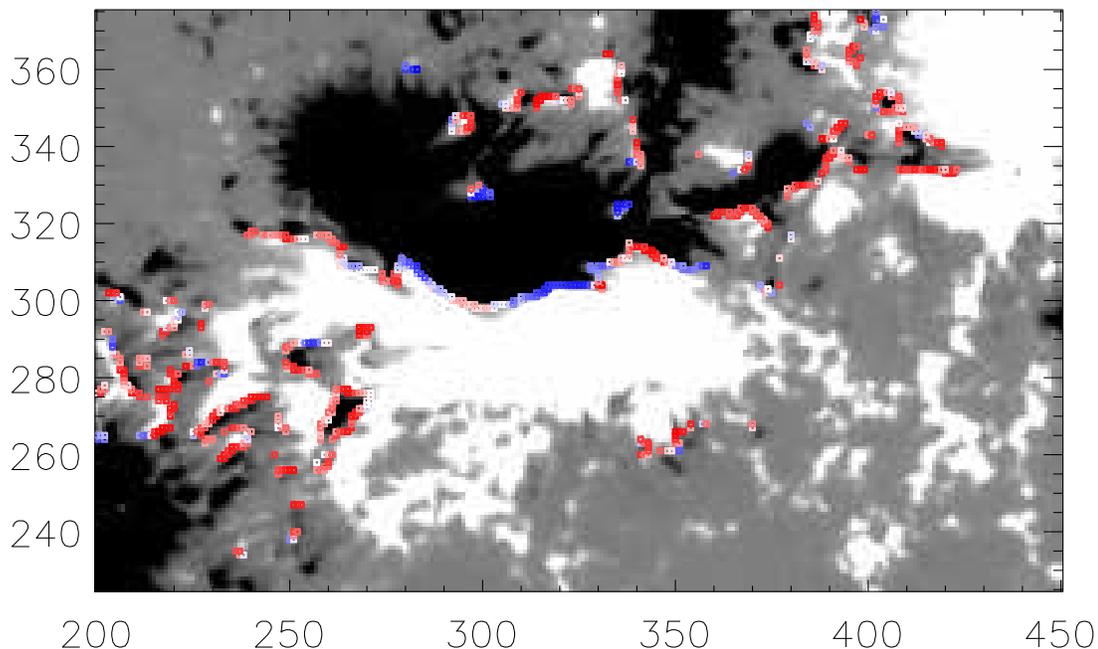,width=7.0in}}
\caption{\footnotesize \textsl{Grayscale image shows a zoomed-in view
    of $B_{\rm LOS}$, with PIL pixels superimposed and color-coded by
    (possibly biased) Doppler velocity $v_{\rm LOS}'$.  Grayscale
    saturation is set to 100 Mx cm$^{-2}$, and the velocity saturation
    is set to $\pm$ 250 m sec$^{-1}$.  Note the predominance of
    redhsifted PILs.}}
\label{fig:find_pils}
\end{figure}

The first step in our approach is to use an automated algorithm to
identify PILs in LOS magnetograms.  Our procedure
creates masks of each polarity for pixels with $|B_{\rm LOS}|$ above a
field strength threshold of $B_{\rm thr}$, dilates each mask by one
pixel, and finds all areas where the dilated masks overlap (see
\citealt{Schrijver2007, Welsch2008b}).  The threshold field is the
only input parameter in PIL identification procedure; here, we use
$B_{\rm thr} = 50$ Mx cm$^{-2}$, ensuring that we identify relatively
strong fields in close proximity.  This approach identifies structures
1--3 pixels wide, which we erode into single-pixel-width lines with
IDL's \texttt{morph\_thin.pro} procedure.  We define the resulting
pixels to be PIL pixels.  In Figure \ref{fig:find_pils}, we show
identified PIL pixels, color-coded by (possibly biased) Doppler
velocity $v_{\rm LOS}'$, in a close-up view of the LOS magnetic field
at the time step when AR 11158 was closest to disk center.

In this sub-region of the full active region, $N_{\rm PIL}= 98$ PILs
were identified.  Most of the Doppler signals along PILs appear
redshifted.  These are almost certainly pseudo-redshifts, arising from
the Doppler zero-point being defined to be blue-ward of the true rest
wavelength for photospheric plasma with no LOS velocity.  The velocity
zero point for HMI is derived from the median of Doppler velocities
within 90\% of the solar radius over 24 hours (Y. Liu \& S. Couvidat,
private communication, 2011).  This implies contributions from
quiet-Sun (and therefore strongly convecting) plasma dominate the
median, and probably bias the zero-point.

\subsection{Flux Changes Near PILs}
\label{subsec:neighborhoods}

Having identified PILs at a given time $t_0$, our second step is to
estimate the set of changes of LOS flux $\{ \Delta \Phi_{\rm LOS}\}$
near all PILs over the $\Delta t = 24$ minutes between the two LOS
magnetograms observed $\pm 12$ minutes from the magnetogram at $t_0$.
To do so, for each PIL we create a binary map of that PIL's pixels,
then apply IDL's \texttt{dilate.pro} procedure with a $(d \times d)$
matrix of 1's to the binary map to define that PIL's ``neighborhood
mask.''  The dilation parameter $d$ is the only other free parameter
in our method.  While granular flows have speeds on the order of $\sim
1$ km sec$^{-1}$, such flows are short-lived, with lifetimes of $\sim
5$ min.  Over $\Delta t = 24$ min, therefore, displacements of fluxes
tend to be consistent with lower averaged speeds, on the order of
$\sim 0.5$ km sec$^{-1}$ or less (see, e.g., Fig. 13 of
\citealt{Welsch2012}).  Given the HMI pixel size $\Delta x \simeq 0.5
\arcsec \simeq 360$ km near disk center, magnetic flux should not move
by more than about two pixels over $\Delta t = 24$ min. Here, we use
$d = 7$ --- that is, dilation by $\pm 3$ pixels in all directions ---
although we have explored other choices (see below).
(As \citet{Demoulin2003} have pointed out, emerging fields that are
strongly inclined with respect to the LOS could have very high
apparent speeds, implying displacements that would exceed our 3-pixel
dilation.  Excessive dilation, however, risks incorporating changes in
flux that are unrelated to emergence / submergence in PIL
neighborhoods.)
Next, we multiply the two co-aligned LOS magnetograms from $\pm 12$
minutes by this neighborhood mask and sum the unsigned flux in each
product, then difference these summed fluxes.  Because this counts LOS
flux changes in {\em both} polarities, we divide the result by two to
estimate $\Delta \Phi_{\rm LOS}$ over $\Delta t = 24$ minutes.

It should be noted that converging or diverging horizontal flows
unrelated to emergence/ submergence at each PIL can transport
``background'' flux into the neighborhood of a PIL, and near-PIL flux
out of this neighborhood, in some cases dominating our estimates of
the change in LOS flux due to emergence or submergence along the PIL.
Hence, the computed rate of change in LOS flux along an individual PIL
is subject to large errors.  Identifying collections of like-polarity
pixels as ``features'' and tracking their evolution
\citep{DeForest2007, Welsch2011} might be one way to better account
for fluctuations in LOS flux due to flux concentration / dispersal
\citep{Lamb2008}. Assuming the contributions from converging and
diverging flows cancel in the aggregate over many PILs, the
statistical properties of changes in LOS fluxes along a set of PILs
are more robust to such errors, and therefore more useful for our
purposes.

Our third step is to compute, for each PIL, the corresponding expected
flux change from the (assumed biased) Doppler signal, $\Delta
\Phi_{vBL}'$, and magnetic length, $\overline{BL}$.  Since the expressions
derived for these quantities (equations \ref{eqn:ideal_equiv_fd} and
\ref{eqn:maglen}) in \S \ref{sec:theory} refer to the component of the
transverse magnetic field perpendicular to the PIL, we investigated the 
angle of the transverse magnetic field along PILs in our dataset.
\begin{figure}[!htb] 
  \centerline{\psfig{figure=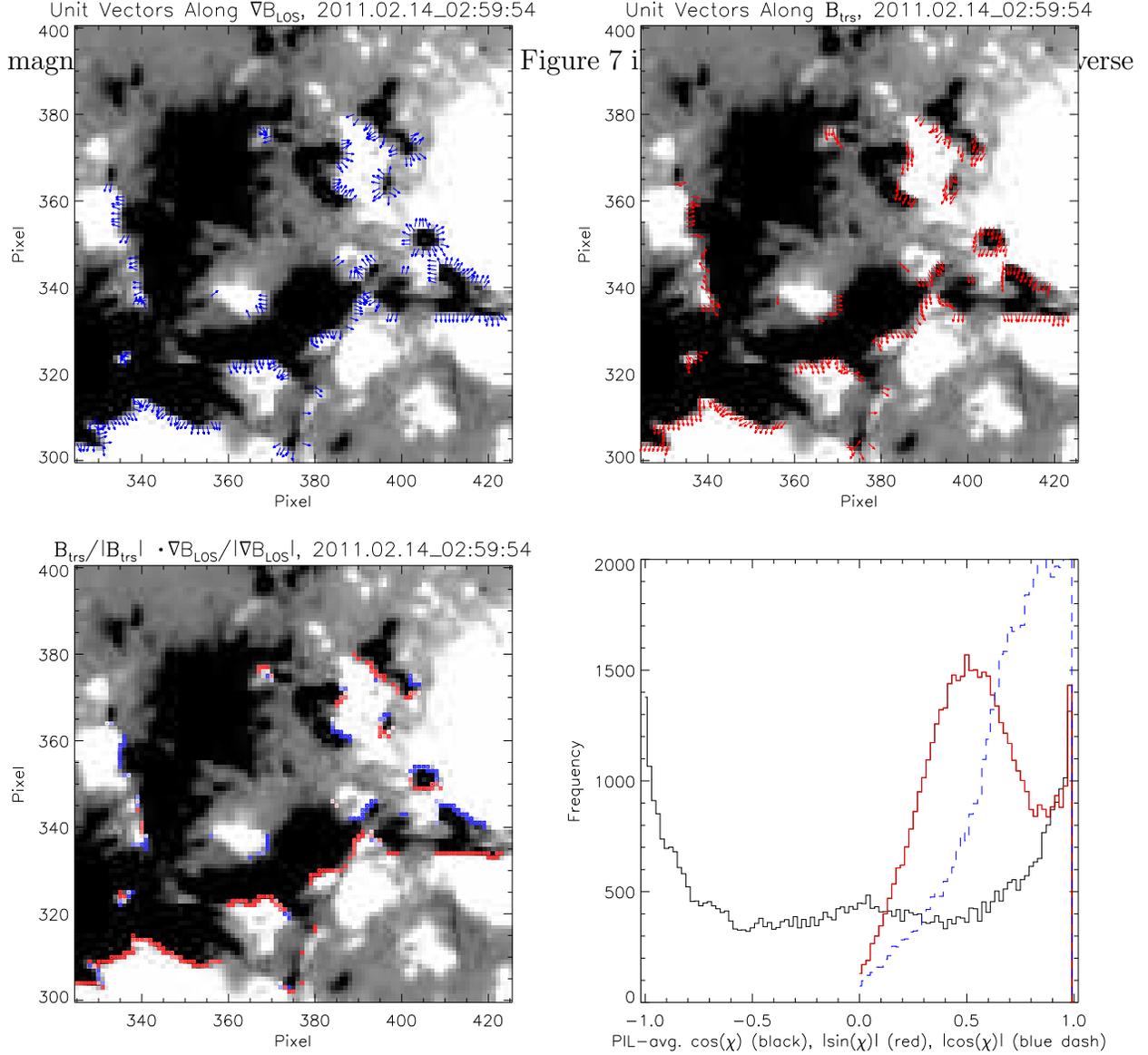,width=6.5in}}
\vspace{0.5in}
\caption{\footnotesize \textsl{Top left: Unit vectors normal
    $\hatx_{\perp}$ to PILs, along the gradient in $B_{\rm LOS}$ along
    each PIL in a small subregion of the full FOV.  Grayscale
    background shows $B_{\rm LOS}$, with saturation at $\pm 100$ Mx
    cm$^{-2}$. Top right: Unit vectors $\hatb_{\rm trs}$ parallel to
    $\bvec_{\rm trs}$ along PILs in the same FOV.  Bottom left:
    Color-coded dot product of $\hatx_{\perp} \cdot \hatb_{\rm trs}
    \equiv \cos \chi$, with red positive (``inverse'' polarity) and
    saturation set to 0.866, corresponding to 30$^\circ$.  The
    component of $\bvec_{\rm trs}$ perpendicular to each PIL is
    homogeneous along many PILs.  Bottom right: Distributions of
    PIL-averaged $\cos \chi$ (black), $| \sin \chi|$ (red), and $|\cos
    \chi|$ (blue), where $\chi$ is the angle between $\hatx_{\perp}$
    and $\hatb_{\rm trs}$.  Fields tend to point across (not along)
    most PILs.}}
\label{fig:pilvects}
\end{figure} 
Figure \ref{fig:pilvects} illustrates the geometry of the transverse
field along several PILs.  The upper-left panel shows unit vectors
$\hatx_{\perp}$ perpendicular to the PIL, determined from the gradient
of $B_{\rm LOS}$ evaluated at PIL pixels.  The upper-right panel shows
unit vectors $\hatb_{\rm trs}$ along $\bvec_{\rm trs}$, evaluated at
PIL pixels.  Note the predominance of unit vectors $\hatb_{\rm trs}$
pointing toward the bottom of the FOV: this coherence might be
spurious, and could arise if the minimization of currents by the
ambiguity resolution algorithm introduces artificial, large-scale
correlations in azimuths.  Color coding in the bottom-left panel shows
the cosine of the angle $\chi$ between these unit vectors --- their
dot product, $\hatx_{\perp} \cdot \hatb_{\rm trs}$ --- with red
positive, and the saturation level set to $\pm 0.866$.  
(The sign of the angle $\chi$ is defined such that the angle between
$\hatx_\perp$ and $\hatx_\parallel$ is +90$^\circ$; see equation
\ref{eqn:hatx}.)
Fields along PILs evidently exhibit components in both the ``normal''
(positive-to-negative; blue) and ``inverse'' (negative-to-positive; red)
directions \citep{Martens2001} perpendicular to the PIL.  
For each PIL, the mean (PIL-averaged) signed and unsigned $\cos(\chi)$
and unsigned $\sin(\chi)$ can be computed; the bottom-right panel
shows the distributions of these averages for all PILs in our dataset.
The peaks at -1 and 1 for the signed $\cos(\chi)$ and around 0.5 for
$|\sin(\chi)|$ imply that the average field of many PILs: (i) is
spatially coherent; and (ii) tends to point primarily across (not
along) the PIL.

As with changes in LOS flux near PILs --- which can arise from
converging or diverging flows unrelated to emergence / submergence ---
some component of Doppler velocities along identified PILs might be
unrelated to emergence / submergence. Slight inaccuracies in PIL
identification, for instance, could lead us to improperly include
flows along the magnetic field (siphon flows) in our flux transport
rates.  Also, filling factors --- due to scattered light within the
HMI instrument, or imaged emission from unresolved, unmagnetized
plasma, or both --- could bias the estimated flux transport rate in a
given pixel
(and contributions from either would plausibly tend to be blue-shifted).
Hence, as with changes in LOS flux near PILs, the inferred rate of
transport of transverse flux along an individual PIL is subject to
large errors. Again, the statistical properties of flux transport
rates along the aggregated collection PILs should be more robust to
error, and therefore more useful for quantifying the pseudo-redshift.

We have also investigated making the simplifying replacement $B_\perp
\rightarrow B_{\rm trs}$ when computing $\Delta \Phi_{vBL}$ and
$\overline{BL}$.  Physically, one can justify this approximation by
noting that the footpoints of fields with components tangent to the
PIL must still thread the photosphere somewhere, if not directly
across the PIL, as would be the case for no magnetic component tangent
to the PIL.  We will show results computed both ways, using either
$B_\perp$ or $B_{\rm trs}$ at each PIL pixel, but unless otherwise
stated, results shown below are derived using $B_{\rm trs}$.  As will be seen,
this did not drastically change the estimates of bias velocities
$v_0$, although quantities derived using $B_\perp$ were substantially
noisier.  We also note that using $B_{\rm trs}$ makes the bias
estimation technique independent of ambiguity resolution.

\subsection{Estimation of Bias Velocities}
\label{subsec:get_vbias}

Finally, we can estimate any bias velocity present.  We first
difference $\Delta \Phi_{\rm LOS} - \Delta \Phi_{vBL}'$ on each PIL to
compute the set of bias fluxes $\{ \Delta \Phi_{\rm bias} \}$ from all
PILs.  We must then estimate the coefficient $v_0$ from the ratios of
$\{ \Delta \Phi_{\rm bias} \}$ to $\{ \Delta t \overline{BL} \}$.

In the data, errors are present in both measured quantities.  Before
estimates of typical uncertainties were published by the HMI team, we
adopted uniform uncertainties of 20 m/s for the Doppler velocities, 25
Mx cm$^{-2}$ for $B_{\rm LOS}$, and 90 Mx cm$^{-2}$ for $B_{\rm trs}$.
This value for the Doppler uncertainty is consistent with expected
near-disk-center noise levels reported by \cite{Schou2012}.  The value
for noise in $B_{\rm LOS}$ is significantly larger than the HMI Team's
since-published estimate the uncertainty of $\sim 6$ Mx cm$^{-2}$ in
720-second data (e.g., Liu et al. 2012a), \nocite{Liu2012a} while
the value for $B_{\rm trs}$ is probably closer to the HMI Team's
estimate of noise in the transverse component on the order of $10^2$
Mx cm$^{-2}$ \citep{Sun2012}.

We estimate $v_0$ and its uncertainty by analyzing the 
the set of ratios $\{ \Delta \Phi_{\rm bias}/(\Delta
t \overline{BL}) \}$ over a subset of identified PILs.  Based upon
both our prior knowledge that the convective blueshift biases the
median-derived estimate of the rest wavelength in HMI Doppler data blue-ward,
and the predominance of redshifted PILs in Figure \ref{fig:find_pils},
we expect that the biased PIL velocities are {\em more red} than the
true velocities. 
Recalling that we use the astrophysical convention that redshifts
correspond to {\em positive} velocities with respect to the observer,
the bias velocity $v_0$ should be {\em positive}.

Not all bias fluxes, however, are consistent with a positive bias
velocity.  This is not obvious from equation (\ref{eqn:los_vbl_bias}),
since it deals with absolute values.  Consequently, we now consider
the different possibilities for the relative sizes of $\Delta
\Phi_{\rm LOS}, \Delta \Phi_{vBL}'$ and $\Delta \Phi_{\rm bias}$.  A
positive $v_0$ is consistent with PILs that obey either
\be \Delta \Phi_{vBL}'  > |\Delta \Phi_{\rm LOS}| 
\label{eqn:good_pils1} \ee
or
\be -|\Delta \Phi_{\rm LOS}| < \Delta \Phi_{vBL}' < 0 
\label{eqn:good_pils2} ~, \ee
since the correction to $\Delta \Phi_{vBL}'$ is $-\Delta \Phi_{\rm
  bias}' = -v_0 \Delta t \overline{BL}$.
For PILs that obey either
\be |\Delta \Phi_{\rm LOS}| > \Delta \Phi_{vBL}' > 0 
\label{eqn:bad_pils1} \ee
or
\be \Delta \Phi_{vBL}' < -|\Delta \Phi_{\rm LOS}| < 0 
\label{eqn:bad_pils2} ~,  \ee
however, $v_0$ would have to be {\em negative} to improve agreement
between the flux changes.
%
%

What fraction of PILs are consistent with a positive bias velocity
$v_0$?  Of the $\sim$50,000 PILs identified in the full dataset (with
$B_{\rm thr} = 50$ Mx cm$^{-2}$), 73.8\% and 5.7\% were consistent
with equation (\ref{eqn:good_pils1}) and (\ref{eqn:good_pils2}),
respectively, and 13.9\% and 6.6\% obeyed equations
(\ref{eqn:bad_pils1}) and (\ref{eqn:bad_pils2}), respectively.
Consequently, $\sim$80\% of PILs are consistent with a positive bias
velocity.
%
%

\begin{figure}[!h] 
%
  \centerline{\psfig{figure=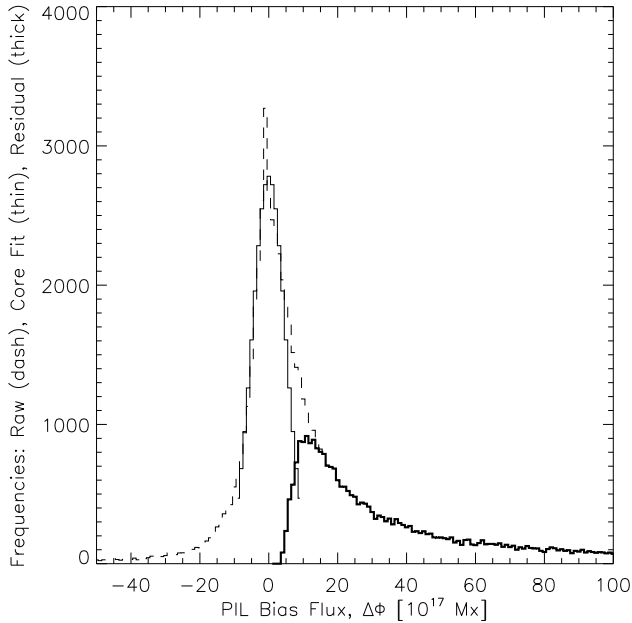,width=3.5in,clip=true}}
\caption{\footnotesize \textsl{Dashed line shows the total
    distribution of bias fluxes for all PILs at all time steps, which
    is clearly skewed toward the positive.  Mirroring the negative
    bias flux values across zero and fitting the result with a
    Gaussian (thin solid) enables an empirical estimate of noise in
    the bias flux from the fitted width, $3.75 \times 10^{17}$
    Mx. Subtracting the fit from the total distribution of bias fluxes
    leaves the residual distribution (thick solid) of bias fluxes
    above the noise level.}}
\label{fig:noise_in_bias}
\end{figure}

In Figure \ref{fig:noise_in_bias}, we show the distribution of bias
fluxes for all PILs at all time steps.  (In any single time step,
there are $\sim 140$ PILs in our FOV, too few to form a continuous
distribution.)  The distribution is strongly skewed toward positive
bias fluxes.  We consider PILs matching either (\ref{eqn:bad_pils1})
or (\ref{eqn:bad_pils2}) to be pathological, which we attribute to
``noise,'' primarily systematic errors in our estimates of the fluxes.
(Errors from our assumed uncertainties in the underlying magnetic and
Doppler data would be $3 \times 10^{16}$ Mx per pixel or less.)  

This distribution of bias fluxes can be used to quantify uncertainties
in the bias flux.  
Mirroring the negative bias flux values across zero and fitting the
result with a Gaussian enables an empirical estimate of noise in the
bias flux from the fitted width, $3.75 \times 10^{17}$ Mx.  Following
the same procedure but using $B_{\perp}$ instead of $B_{\rm trs}$ to
compute $\Delta \Phi_{vBL}'$ results in a fitted width of $5 \times
10^{17}$ Mx, reflecting greater uncertainty in estimates of this flux.
These define thresholds in the bias flux: estimates of bias fluxes
smaller than these levels fall within our (systematic) uncertainties.

Subtracting the fit from the total distribution of bias fluxes leaves
the residual distribution of bias fluxes above our uncertainty
level.  At each step, we computed separate estimates of the bias
velocity from both the full set of PILs and just the subset of PILs
with bias fluxes more positive than these estimates of the threshold
bias fluxes.  Not surprisingly, estimates of $v_0$ using only PILs
with bias fluxes above these thresholds --- which we refer to as the
``high-bias'' subset of PILs --- were substantially higher (as shown
below) than estimates derived using all PILs.  We view use of all PILs
as more conservative, so results shown below are derived from all
PILs, unless explicitly stated otherwise.

\section{Results \& Analysis}
\label{sec:results}

\begin{figure}[!h] 
%
  \centerline{\psfig{figure=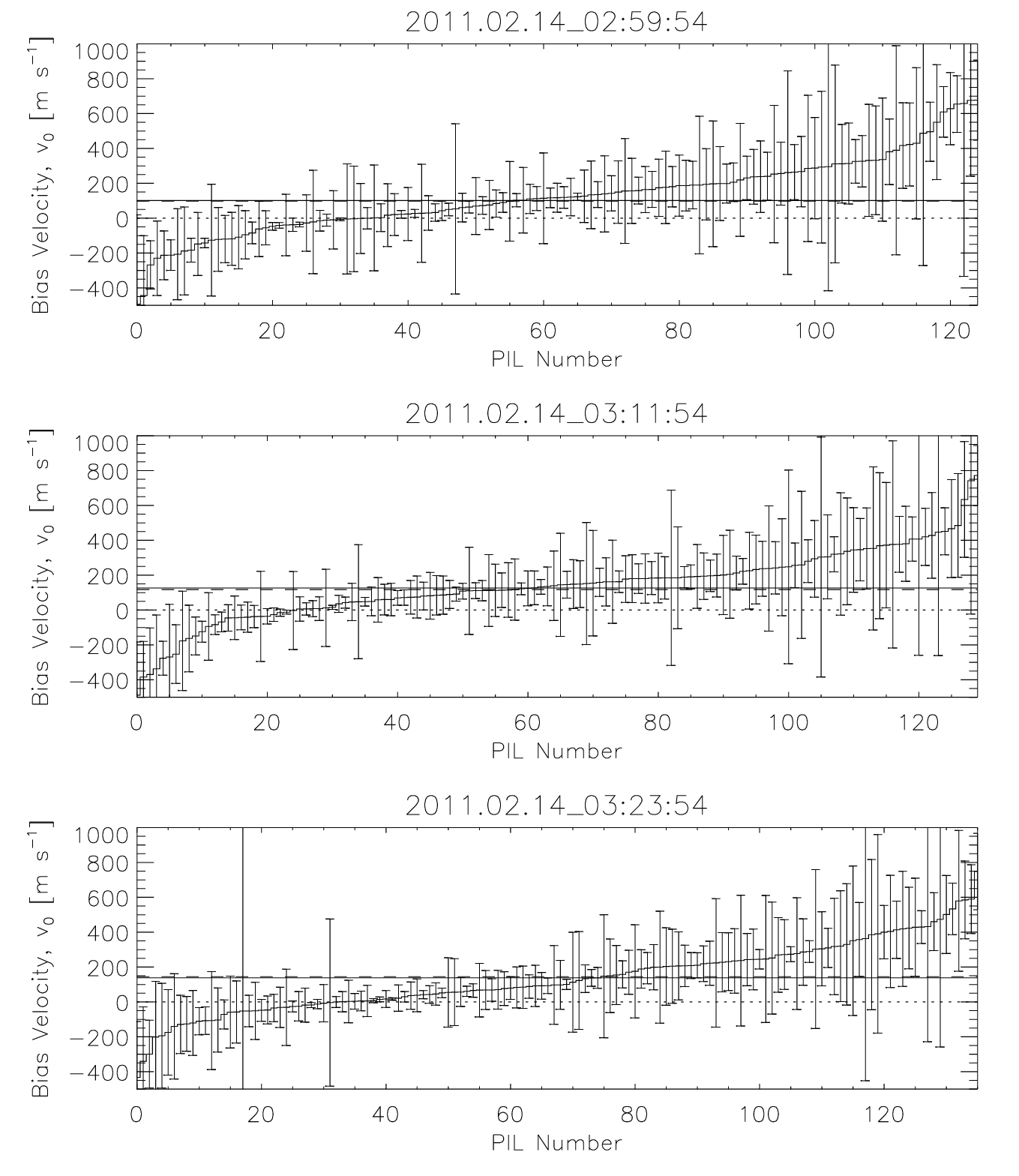,width=6.5in,clip=true}}
\caption{\footnotesize \textsl{Sorted values of the set of bias
    velocities, $\{v_0\} =\{ \Delta \Phi_{\rm bias}/(\Delta t
    \overline{BL}) \}$ from all PILs, (in m/s) and estimated errors
    for each PIL (with the astrophysical convention of positive
    redshift), from three successive (and partially overlapping)
    magnetogram triplets. The lower thin dotted line shows the
    (uncalibrated) zero Doppler velocity.  Solid and dashed horizontal
    lines show average and median bias velocities.  The magnitudes of
    the estimated bias velocities are
    $101 \pm 17$ m/s, $126 \pm 18$ m/s, and $138 \pm 16$ m/s.  
    Only about $\sim 45$\% of error bars are consistent with each
    average velocity, suggesting our error estimates are too low, or
    the ideal assumption is invalid. Nonetheless, our results are
    consistent with the predominance of redshifts in Figure
    \ref{fig:find_pils}, and indicate pseudo-redshifts are present.
}}
\label{fig:v0}
\end{figure}

In Figure \ref{fig:v0}, we show sorted values of the set $\{ v_0 \} =
\{ \Delta \Phi_{\rm bias}/(\Delta t \overline{BL}) \}$ from all PILs,
scaled to units of m/s, derived from three successive magnetogram
triplets when AR 11158 was close to the central meridian.  It is
clear that most values fall well above zero (the lower, dotted line in
each plot; recall that we define redshifts as positive).  This is
entirely consistent with the predominance of redshifts in Figure
\ref{fig:find_pils}.  The means and standard errors in the estimates
are
$101 \pm 17$ m/s, $126 \pm 18$ m/s, and $138 \pm 16$ m/s%
, respectively; the means are plotted as horizontal solid lines.
Medians are the dashed horizontal lines in each plot
(indistinguishable from the mean in the bottom plot).  We use the
standard error in the estimate (standard deviation divided by
$\sqrt{N_{\rm PIL}}$) as a measure of uncertainty since each estimate
is a measurement of the same number, $v_0$, although we do not know if
the errors are Gaussian.  Only $\sim$ 45\% of error bars are
consistent with the average $v_0$ in each plot.  This suggests either:
(1) that our error estimates are too low, perhaps due to our neglect
of systematic errors (e.g., in definitions of PIL and their
neighborhoods, and in quantifying magnetic fields along PILs); or (2)
that electric fields along PILs are in some cases inconsistent with
the ideal electric field assumed in equation (\ref{eqn:ideal}).

\begin{figure}[!htb] 
%
  \centerline{\psfig{figure=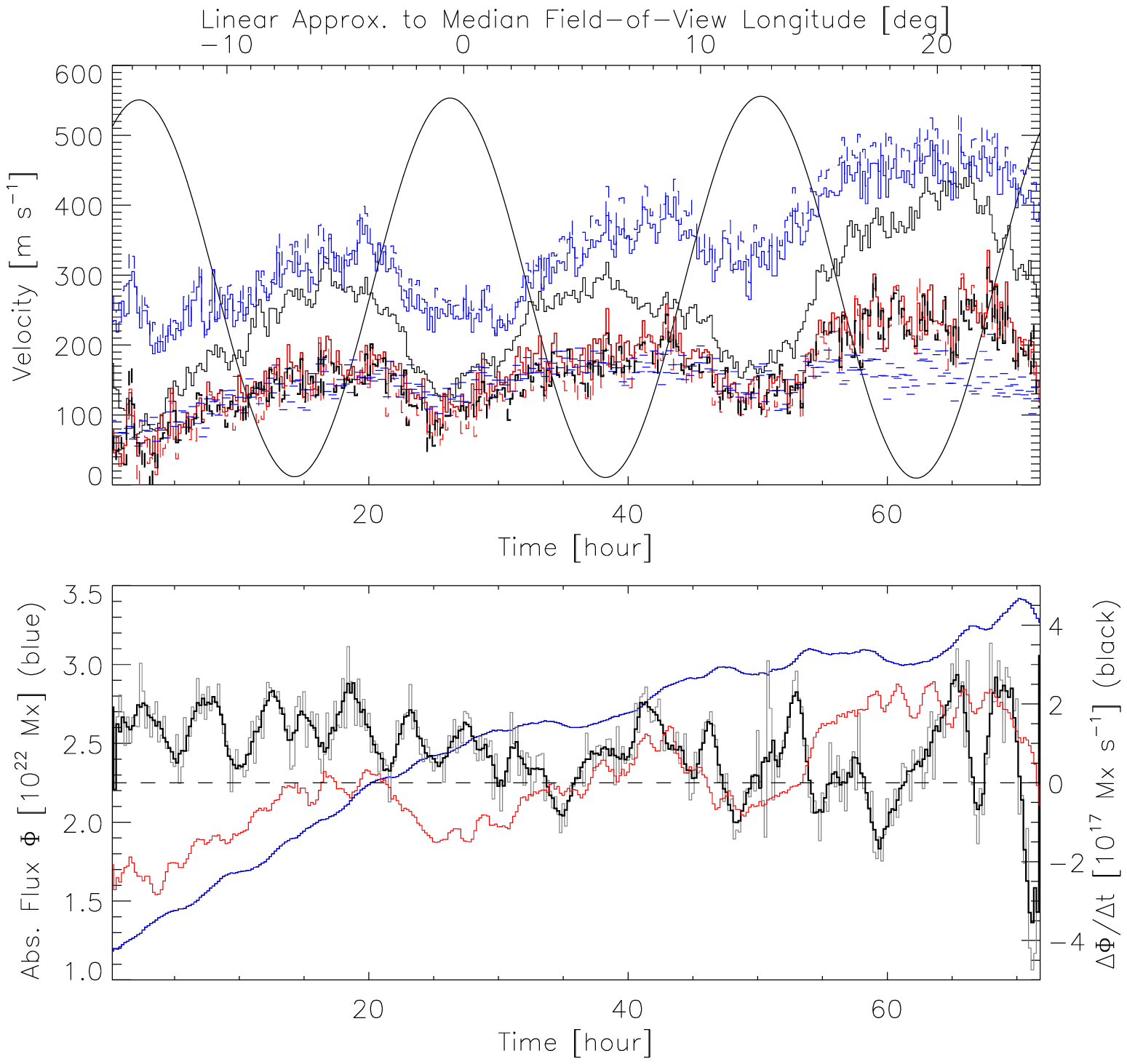,width=5.5in,clip=true}}
\caption{\footnotesize \textsl{Top: Values for the bias velocity $v_0$
    for different approaches and parameters used to determine changes
    in flux near PILs.  The median of all PILs, using $B_{\rm trs}$,
    dilation parameter $d=5$ pixels, and LOS field strength
    significance threshold $B_{\rm thr} = 50$ Mx cm$^{-2}$ gives the
    solid red line.  Error bars, omitted for clarity, are on the order
    of the fluctuations between successive estimates.  With the same
    parameters, but using $B_\perp$, the dashed red line results.  The
    solid and dashed blue lines correspond to estimates using the same
    approaches and parameters as each red line, respectively, but from
    averaging over only high-bias PILs (see text).  The dashed, thick
    black line (which coincides closely with the red lines) shows the
    bias velocity using $B_{\rm trs}$, but $d=7$, keeping $B_{\rm thr}
    = 50$ Mx cm$^{-2}$.  The blue dash marks show the bias velocity
    using $d=5$, with $B_{\rm thr} = 25$ Mx cm$^{-2}$.  The solid
    black jagged line shows an empirical estimate of $v_0$: the median
    of Doppler velocity along all PIL pixels identified with the $50$
    Mx cm$^{-2}$ threshold, which ignores the flux-emergence
    constraint in equation (\ref{eqn:ideal_equiv_fd}).  The smooth,
    sinusoidal solid line shows 1/10th of the radial component of the
    spacecraft Doppler velocity, offset by +250 m s$^{-1}$ to fit on
    the plot.  Bottom: Total unsigned radial flux (blue) and the raw
    and smoothed finite-difference time rate of change in that flux
    (thin gray and thick black, respectively), which is positive when
    new flux is emerging.  Variations in the rate of change of flux
    are weakly correlated with the smoothed, de-meaned, and scaled
    bias velocity estimate (red) derived using all PILs, with $d=5$
    and $B_{\rm thr} = 50$ Mx.}}
\label{fig:bias_vs_time}
\end{figure} 

In the top panel of Figure \ref{fig:bias_vs_time} we plot estimates of
$v_0$ versus time, for different approaches --- using either $B_{\rm
  trs}$ or $B_\perp$, and either all PILs or only high-bias PILs ---
along with different choices for the two parameters in our method, the
threshold LOS field strength $B_{\rm thr}$ (25 or 50 Mx cm$^{-2}$)
that defines significance in identifying PILs, and the dilation
parameter $d$ (5 or 7) that defines the neighborhood near each PIL
over which changes in LOS flux with time are computed.  Estimates made
from the median $v_0$ from all PILs, with $B_{\rm thr} = 50$ Mx
cm$^{-2}$ and $d=5$ using either $B_{\rm trs}$ or $B_\perp$ (red,
solid and dashed, respectively) are very similar.  (We use medians
instead of averages here because some estimates in the full set of
PILs made using $B_\perp$ are pathologically noisy.)  The blue lines
are from averaging only high-bias PILs, using either $B_{\rm trs}$ or
$B_\perp$ (solid and dashed, respectively), and also agree closely.
Using $B_{\rm trs}$, increasing $d$ to 7, with $B_{\rm thr} = 50$ Mx
cm$^{-2}$, gives the thick, black dashed line (which closely follows
the red lines); decreasing $B_{\rm thr}$ to $25$ Mx cm$^{-2}$, with
$d=5$, gives the blue dash marks (which are also close to the red
lines).  Together, these suggest that results are not strongly
affected by varying $d$ slightly, or using $B_{\rm trs}$ versus
$B_\perp$, although estimates of $v_0$ from high-bias-flux PILs are
systematically higher.  We note that the less restrictive threshold
flux density, $B_{\rm thr} = 25$ Mx cm$^{-2}$, resulted in somewhat
lower estimated bias velocities, especially toward the end of the run.
This might be because PILs in weaker-field regions, which should be
more strongly convecting, were included when the lower threshold was
used.

The HMI Team's estimate of the 1$\sigma$ noise in $B_{\rm LOS}$ is
about $6$ Mx cm$^{-2}$, so the $50$ Mx cm$^{-2}$ threshold corresponds
to about 8 standard deviations.  With $B_{\rm thr} = 50$ Mx cm$^{-2}$
and $d=5$, the standard deviation between the $i$-th bias estimate
made using $B_{\rm trs}$, and a 5-step (1-hr.) running average was 18
m/s, near the noise level of the Dopplergrams.  Standard errors in the
mean (SEM) were computed using uncertainties assumed above; the
average SEM for this series was 22 m s$^{-1}$.  The consistency of the
$v_0$ estimates on hour-long time scales is evidence of robustness in
our estimates.  Variation of $v_0$ on time scales longer than few
hours is evidence that calibration of the Doppler shifts in time is
necessary, as the estimated bias velocity is not constant.  For
comparison with the phase of SDO's orbit, we also plot the radial
component of spacecraft velocity, rescaled and shifted in the vertical
direction (but not in time), in the sinusoidal, black, solid line.

The solid, black, jagged line shows an empirical estimate of $v_0$:
the median of Doppler velocities along all PIL pixels at each step,
identified with the $50$ Mx cm$^{-2}$ threshold.  This estimate is
relatively simple to determine: one only needs to identify PILs of the
LOS field, then take the median Doppler velocity on all PIL pixels.
This approach, however, ignores the flux-matching constraint in
equation (\ref{eqn:ideal_equiv_fd}), and therefore could be biased by
a strong episode of flux emergence (or submergence).

A trend in each curve with longitude (upper scale, Figure
\ref{fig:bias_vs_time}) can be noted, of about 5 m s$^{-1}$ deg$^{-1}$
for the curves shown.  In terms of an error in our assumed rotation
rate, this corresponds to $\sim 0.45 \mu$rad s$^{-1}$, a significant
fraction of the $A$ coefficient in equation (\ref{eqn:snod}).  We note
that accurate compensation for the rotation rate prior to applying our
method is not required: it can also be applied even if the rotational
Doppler shift has not been removed, although the derived bias velocity
will then include both rotational and convective shifts.

A more likely explanation for the trend is contamination by Evershed
flows on ``false'' PILs in the LOS field of limb-ward penumbrae.  Such
artifacts might be ameliorated by only including bias velocity
estimates from LOS PILs that are near radial-field PILs.

All of our estimates of the bias velocity $v_0$ are positive, although
there is a significant spread in the estimates, which range from 50 --
500 m s$^{-1}$, with variations in both time (and therefore longitude)
and from method and parameter selection.  We will compare the physical
implications of the differing estimates below.

\begin{figure}[!htb] 
%
  \centerline{\psfig{figure=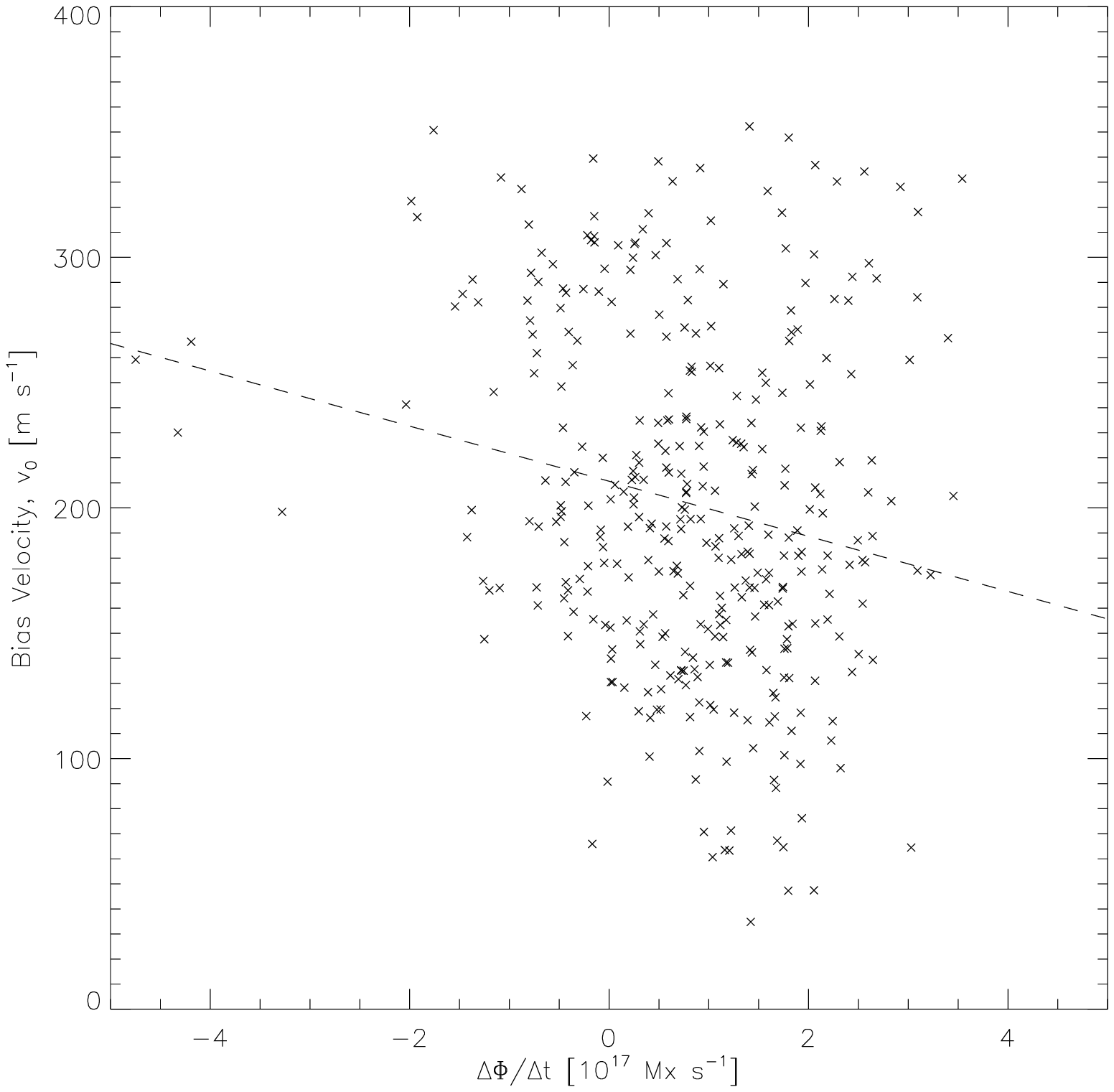,width=3.5in,clip=true}}
\caption{\footnotesize \textsl{A scatter plot of the bias velocity as
    a function of the rate of change of magnetic flux, which are
    weakly correlated. A linear fit (dashed line) with slope 11 m
    s$^{-1}$ (10$^{17}$ Mx s$^{-1}$)$^{-1}$ is overplotted.}}
\label{fig:vbias_dbdt}
\end{figure}

%
%
Could an episode of strong flux emergence somehow influence our
estimate of the bias velocity?  In principle, the flux-matching
constraint in equation (\ref{eqn:ideal_equiv_fd}) should make our
approach insensitive to the rate of flux increase or decrease: changes
in LOS flux and the transport of flux along the LOS should be
consistent, regardless of emergence / submergence rates.  Nonetheless,
could systematic errors in our approach make our calibration method
susceptible to bias during episodes of strong emergence?  We
investigate this possibility in Figure \ref{fig:bias_vs_time}'s bottom
panel, which shows the total unsigned radial flux in pixels with
absolute radial flux density above 50 Mx cm$^{-2}$, and the raw and
smoothed (using a five-step boxcar average) finite-difference time
rate of change in that flux, which is positive when flux is emerging.
The smoothed, de-meaned, and rescaled bias velocity derived from all
PILs with $B_{\rm thr} = 50$ Mx cm$^{-2}$ and $d=5$ is also plotted
(red line).  Although variations in the rate of change of flux do not
appear to strongly influence the bias velocity estimates, we find a
weak correlation ($\sim 0.2$) between these time series.  A scatter
plot, in Figure \ref{fig:vbias_dbdt}, confirms that the dependence is
weak.  Nonetheless, we ran a total-least squares linear fit to
quantify the dependence, which gives a slope of -11 m s$^{-1}$
(10$^{17}$ Mx s$^{-1}$)$^{-1}$, implying that typical fluctuations of
$\Delta \Phi/\Delta t$, on the order of $2 \times 10^{17}$ Mx, should
influence the estimated bias velocity by about 20 m s$^{-1}$, the
noise level of the Dopplergrams themselves.  As noted above, this
correlation might arise indirectly from some systematic aspect of our
method. An alternative explanation for this correlation is the known
instrumental correlation between the spacecraft's periodic Doppler
motion and the observed periodicities in HMI's LOS field strengths
(Liu et al. 2012a).  \nocite{Liu2012a} The correlation between the
median Doppler velocity on PILs and the rate of change of flux is
similarly weak, with a fitted slope of -9 m s$^{-1}$.

\begin{figure}[!htb] 
%
  \centerline{\psfig{figure=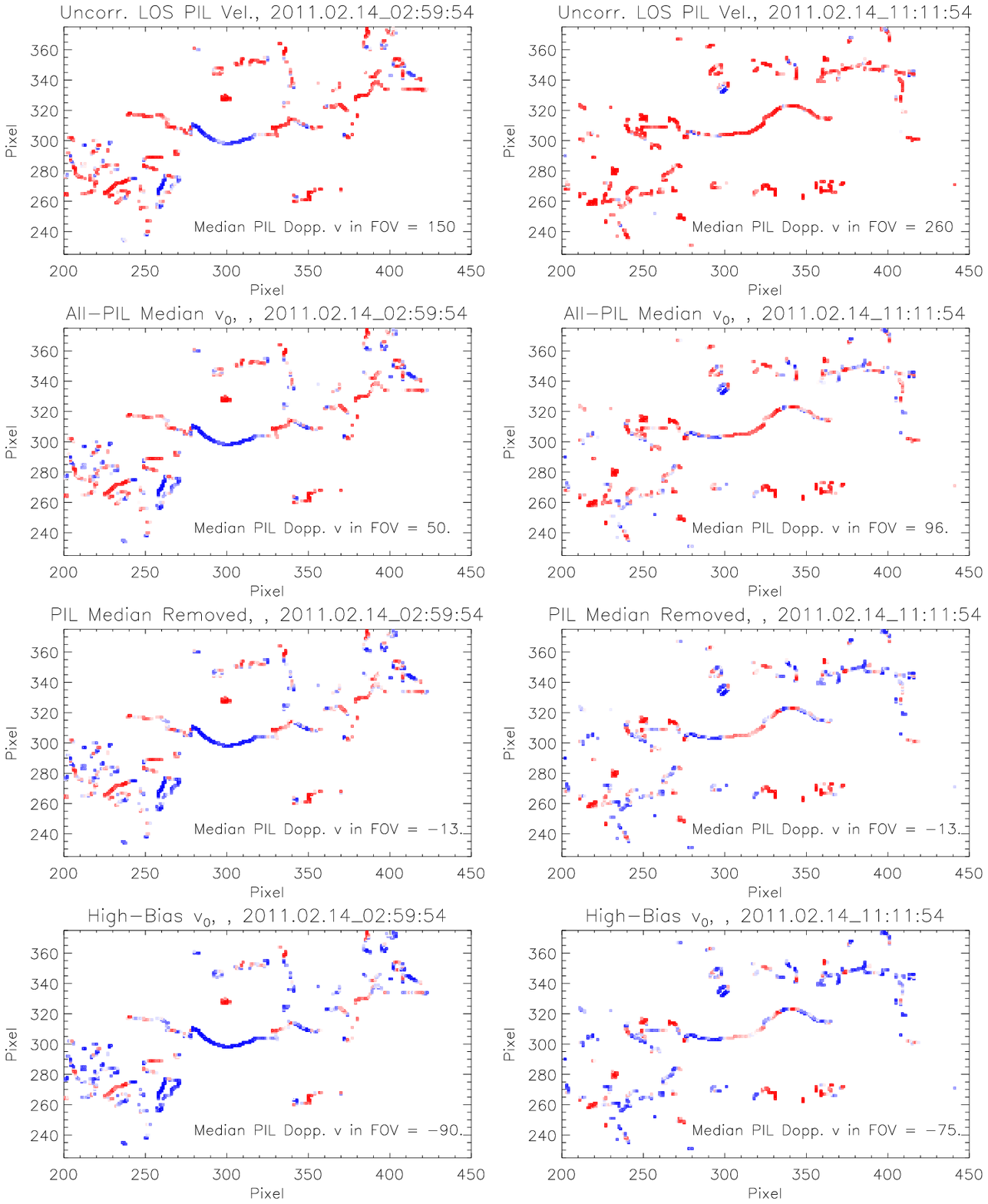,width=6.5in,clip=true}}
\caption{\footnotesize \textsl{Examples of Doppler shifts along
    identified PILs in two HMI LOS magnetograms (separated by about
    eight hours), both without correction for the bias velocity (top
    row) and corrected using different estimates of $v_0$.  Saturation
    in the color map is set at $\pm 250$ m/s in all panels.  Second
    row: Doppler velocities corrected using $v_0$ derived by averaging
    over all PILs. Third row: Doppler velocities resulting by removing 
    the median velocity on all PIL pixels. Fourth row: Doppler
    velocities corrected using $v_0$ derived by averaging over
    high-bias PILs, those with bias fluxes above $3.75 \times 10^{17}$
    Mx.  Almost all PILs in the top row appear redshifted, but redshifts
    and blueshifts are more evenly balanced along PILs in the lower
    rows.}}
\label{fig:corrected_pils}
\end{figure} 

In Figure \ref{fig:corrected_pils}, we show Doppler shifts along
identified PILs in two magnetograms, recorded about eight hours apart
(left and right columns), both uncorrected (top row) and corrected by
subtraction of different estimates of $v_0$ made in different ways:
the average $v_0$ from all PILs (second row); the median Doppler
velocity of all PIL pixels (third row); and the high-bias estimate of
$v_0$, using only PILs with bias fluxes above $3.75 \times 10^{17}$ Mx
(fourth row).  While redshifts predominate along PILs in the top row,
Doppler shifts along PILs in the bottom rows are more evenly balanced
between red- and blueshifts.  

Our estimates of the bias velocity fall in the range of a few hundred
m s$^{-1}$.  \citet{Asplund2003} used 3D, radiative MHD simulations of
magnetoconvection to study line profiles for several spectral lines of
iron, including some similar to HMI's Fe I 6173 \AA \, line in
wavelength.  They find convective blue shifts similar in magnitude to
our estimates of the pseudo-redshift (around 300 -- 500 m/s), with
uncertainties (50 -- 100 m/s), similar to ours for $v_0$.  Only $v_0$
estimates from high-bias PILs fall in the range found by
\citet{Asplund2003}, but it should be noted that their estimate of the
convective blueshift is with respect to the true rest frame, but our
estimate of the bias velocity is with respect to the HMI Team's
estimate of the rest wavelength, derived from the median of 24 hours
of HMI's measured Doppler velocities within 90\% of the solar radius.

\begin{figure}[!htb] 
%
  \centerline{\psfig{figure=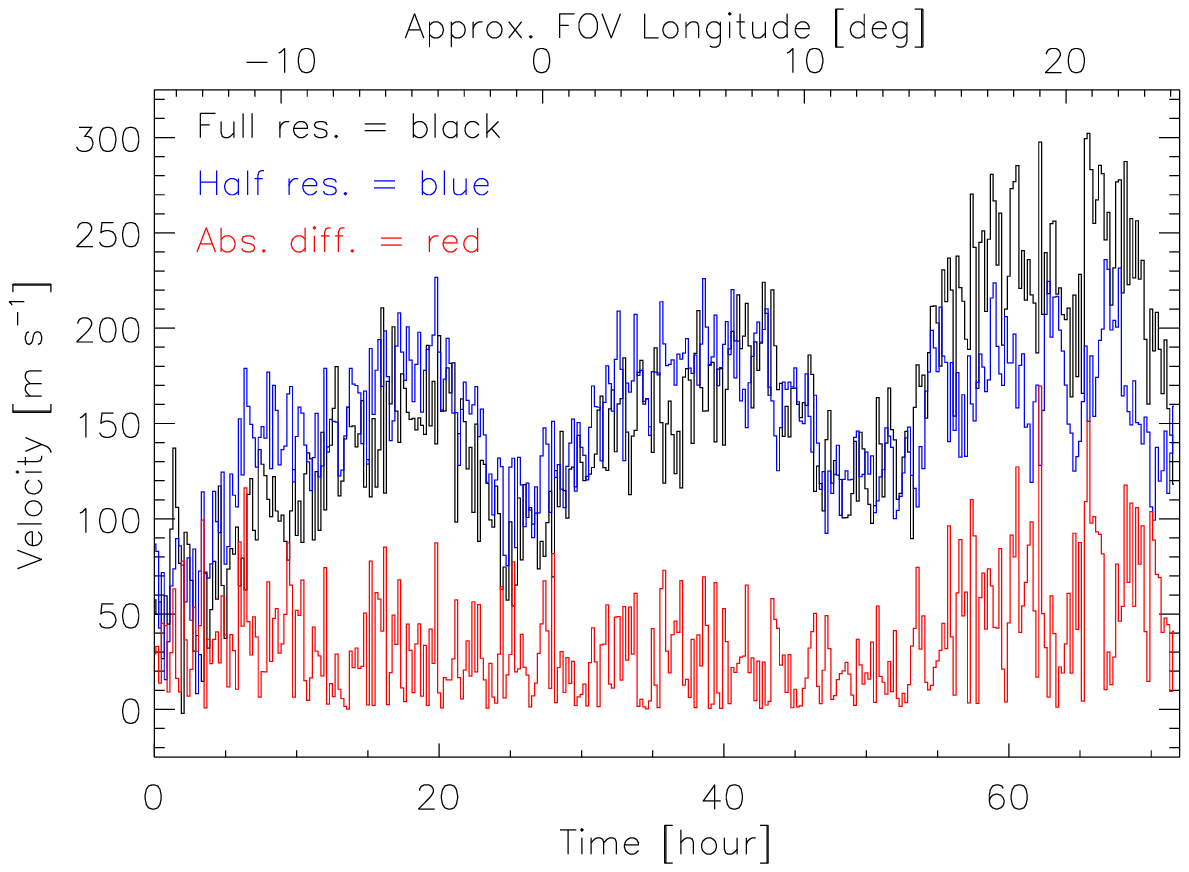,width=5.5in,clip=true}}
\caption{\footnotesize \textsl{Bias velocities from full-resolution
    (black) and $(2 \times 2)$-binned (blue) data are significantly
    correlated, with rank-order and linear correlation coefficients
    near 0.6.  The absolute difference between the two is plotted in
    red; the mean and median differences are 35 m s$^{-1}$ and 29 m
    s$^{-1}$, respectively.}}
\label{fig:halfres}
\end{figure}

We note that the equations used in our approach do depend explicitly
upon spatial resolution: formally, changes in LOS flux should be
consistent with up- or downflows along PILs, regardless of resolution.
In practice, however, diminished spatial resolution could reduce the
observable changes in unsigned LOS flux, since unresolved positive and
negative LOS flux would cancel.  To test the effect of spatial
resolution on the method, we applied it to data in which the
resolution was artificially decreased by binning pixels $(2 \times
2)$.  Figure \ref{fig:halfres} shows the median bias velocity at each
step for each case, with dilation parameters $d = 5$ pixels in both.
The bias velocities were significantly correlated, with rank-order and
linear correlation coefficients near 0.6.  Results were very similar
with $d=3$ and $d=7$ in the rebinned data.  In all cases, the largest
discrepancies occurred when the median FOV longitude was near 20
degrees; rebinning appears, for some reason, to affect results as
central meridian distance increases.
This suggests our approach is robust enough for use with data sets of
different resolution.  We note that median PIL Doppler velocities were
very highly correlated, with with rank-order and linear correlation
coefficients near 0.95, suggesting this estimator of bias velocities
is very robust.

As an aside, we remark that we found Doppler structures along PILs to
persist from one HMI vector magnetogram to the next, i.e., with a
lifetime of at least 720 s.  The lifetimes of patterns of upflows and
downflows along PILs have not yet been studied yet, but bear
investigation.  Autocorrelation of Dopplergrams might be useful for
this purpose, similar to methods used by \citet{Welsch2012}.

\subsection{Physical Processes Along PILs}
\label{subsec:processes}

To investigate which of our estimates for $v_0$ is most reasonable, we
now consider the physical implications of our results in terms of the
upward and downward transport of flux across the photosphere.
Recall that processes that increase photospheric flux are
$\Omega$-loop emergence and $U$-loop submergence, while processes that
remove flux are $\Omega$-loop submergence and $U$-loop emergence.
Without correcting Doppler velocities, essentially all increases and
decreases in flux are attributed to the two submergence processes
(from $U$-loops and $\Omega$-loops, respectively), since nearly all
PILs show only redshifts.  Adjusting the observed Doppler shifts to
compensate for the convective blueshift, however, leads to a different
apportionment between the four possible processes, depending upon the
applied bias velocity: PILs that, on average, show upward transport of
transverse flux and increases (decreases) in LOS flux are ascribed to
$\Omega$-loop ($U$-loop) emergence, while PILs that, on average, show
downward transport of transverse flux and decreases (increases) in LOS
flux are ascribed to $\Omega$-loop ($U$-loop) submergence.

\begin{figure}[!htb] 
%
  \centerline{\psfig{figure=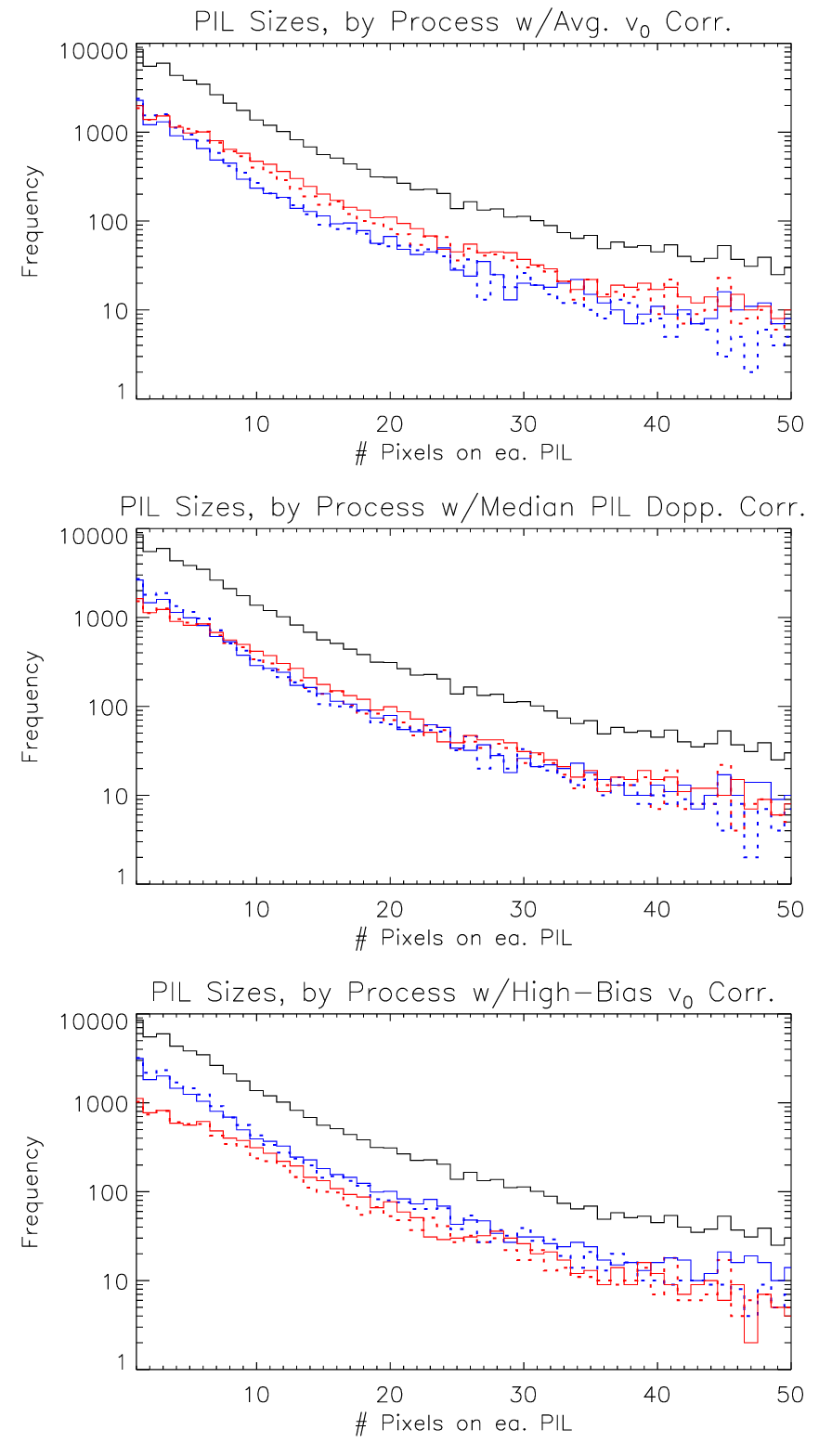,width=3.75in,clip=true}}
\caption{\footnotesize \textsl{Frequencies of emergence and
    submergence processes, as functions of the number of pixels along
    all identified PILs at all time steps, for three possible
    corrections to Doppler velocities: the average estimates of $v_0$,
    derived using $B_{\rm trs}$ with $B_{\rm thr} = 50$ Mx cm$^{-2}$
    and $d=5$ (top); the median of Doppler velocities on pixels in
    identified PILs (middle); and the average high-bias estimates of
    $v_0$, again using $B_{\rm trs}$ with $B_{\rm thr} = 50$ Mx
    cm$^{-2}$ and $d=5$ (bottom).  In all panels, the black curve
    shows the distribution of sizes of all PILs. Processes involving
    blueshifts ($\Omega$-loop and $U$-loop emergence) are plotted in
    blue, while processes involving redshifts ($\Omega$-loop and
    $U$-loop submergence) are plotted in red.  Processes that remove
    flux via cancellation ($U$-loop emergence and $\Omega$-loop
    submergence) are plotted with dashed lines.}}
\label{fig:pilpix_plots}
\end{figure}

In Figure \ref{fig:pilpix_plots}, we show the frequencies of these
processes, as functions of the number of pixels along all identified
PILs at all time steps, for three possible corrections: the average
estimates of $v_0$, derived using $B_{\rm trs}$ with $B_{\rm thr} =
50$ Mx cm$^{-2}$ and $d=5$ (top); the median of Doppler velocities on
pixels in identified PILs (middle); and the average high-bias
estimates of $v_0$, again using $B_{\rm trs}$ with $B_{\rm thr} = 50$
Mx cm$^{-2}$ and $d=5$ (bottom).  In all panels, the black curve shows
the distribution of sizes of all PILs.  Processes involving blueshifts
($\Omega$-loop and $U$-loop emergence) are plotted in blue, while
processes involving redshifts ($\Omega$-loop and $U$-loop submergence)
are plotted in red.  Processes that remove flux via cancellation
($U$-loop emergence and $\Omega$-loop submergence) are plotted with
dashed lines.  Even with the average-$v_0$ correction (top panel),
submergence processes still dominate, contradicting the expectation
that the observed increase in unsigned radial flux of $\sim 2 \times
10^{22}$ Mx over the data sequence occurred via emergence from the
interior.  Removing the median Doppler velocity over all pixels in
identified PILs at each time step gives a more reasonable distribution
of processes (middle panel): emergence and submergence processes are
more closely balanced. But the requirement of a net flux increase
implies that emerging processes should predominate, and this
expectation is only consistent with the high-bias-$v_0$ correction
(bottom panel).

In principle, observations of changes in total unsigned radial flux in
the FOV should be consistent, separately, with estimates of changes in
LOS flux near PILs and the upward / downward transport of transverse
flux.  Our estimates of flux increase by $\Omega$-loop emergence from
changes in LOS flux near PILs are roughly consistent with the net
change in LOS flux over the entire sequence. 
Unfortunately, other components of our flux budgeting near PILs are
inaccurate, in two ways.

First, our estimates of flux removed by cancellation processes
($\Omega$-loop submergence and $U$-loop emergence) are too large (from
$\sim 10^{22}$ Mx to several times this), whether inferred from LOS
changes or rates of transport of transverse flux.
This might be explained if some flux is dispersing near PILs that we
identify as sites of cancellation, instead of submerging/emerging at
these PILs.
As discussed in Section \ref{subsec:neighborhoods}, identifying
magnetogram features and tracking their evolution
\citep{DeForest2007}, or imposing additional constraints (e.g.,
requiring cancellation to persist for multiple time steps, or taking
the flux lost to be the minimum of that among the two polarities;
see \citealt{Welsch2011}) could more accurately quantify canceled flux.
We leave investigation of these approaches for future work.

Second, our estimated rates of transport of transverse flux are
systematically too large compared to changes in LOS flux near PILs,
for all processes.
%
%
As noted in Section \ref{subsec:neighborhoods} above, inaccuracies in
PIL identification could incorporate velocities from siphon flows in
our flux transport rates.  In addition, a filling factor $f < 1$ could
also play a role, since it would affect $|\Delta \Phi_{\rm LOS}|$ and
$|\Delta \Phi_{vBL}|$ differently: while length transverse to the LOS
(two factors of which enter computation of $|\Delta \Phi_{LOS}|$) is
scaled by $\sqrt{f}$, length along the LOS (which enters computation
of $|\Delta \Phi_{vBL}|$) is not.  Hence, if $f < 1$, one would expect
$|\Delta \Phi_{vBL}| > |\Delta \Phi_{\rm LOS}|$.  Given typical
discrepancies in fluxes by factors of 4 -- 6, this would imply filling
factors of $\sim 0.05$.  This would be implausible in umbrae, where
the field is expected to be space-filling.  We analyze PILs, however,
which are, almost as a matter of definition, in mixed-polarity
regions, where the field is expected to be more intermittent.  The
plausibility of a filling factor this small along PILs is unclear.

Another possibility is that flux transport rates are overestimated
because the transport is not ideal, i.e., that the plasma is moving at
a different speed than the flux transport rate.  For instance, if the
flux emergence process were not ideal \citep{Pariat2004}, then the
rate of flux transport derived from Doppler shifts would underestimate
the true emergence rate, since flux would be partially decoupled from
the plasma, allowing flux to move upward independently of the plasma.
While magnetic diffusivity would be manifested in that case by flux
slipping through stationary (or much slower-moving) plasma at a rate
faster than that suggested by ideal transport, the ideal transport
rate is too high in our estimates.  \citet{Chae2008} note that
turbulent diffusivity can act on canceling LOS magnetic fluxes; but
again, we find the rate of loss of LOS field to be too small relative
to ideal transport of horizontal fields, and invoking turbulent
diffusivity would make this discrepancy worse.  One possible
explanation is that unmagnetized plasma, which could be systematically
faster-moving than magnetized plasma, could be contributing
significantly to Doppler signal.
 
To quantitatively characterize cases of apparently non-ideal
evolution, one could employ a simple non-ideal model, assuming a uniform
magnetic diffusivity $\eta$ is present.  This would imply a
corresponding non-ideal electric field parallel to the PIL (in
addition to the ideal electric field along the PIL already used to
estimate $\Delta \Phi_{vBL}$), given by
\be c E_\parallel^{\rm NI} = \eta |\partial_{\rm LOS} B_{\perp} 
- \partial_\perp B_{\rm LOS}| ~, \label{eqn:nonideal} \ee
where, as above, $\hatx_\perp$ locally points across the PIL (see
equation \ref{eqn:hatx}).  For PILs where $|\Delta \Phi_{\rm LOS}| \gg
|\Delta \Phi_{vBL}|$, we could attribute the excess $\Delta \Phi_{\rm
  LOS}$ to $\eta \, \partial_\perp B_{\rm LOS}$ integrated along the
PIL, taking $\partial_{\rm LOS} B_{\perp}$ to be negligible, as it
would be if there were a current sheet along the PIL (e.g.,
\citealt{Litvinenko1999a}).  Values of $\eta$ derived this way could
be compared to the assumed values used by \citet{Litvinenko1999a},
\citet{Chae2002}, and \citet{Litvinenko2007}.  For PILs where $|\Delta
\Phi_{\rm LOS}| \ll |\Delta \Phi_{vBL}|$, we could attribute the
excess $\Delta \Phi_{vBL}$ to overestimation of the transport rate of
transverse flux along the LOS, due to slippage of transverse flux
through the moving plasma in the presence of diffusivity.  Because
$\partial_{\rm LOS} B_{\perp}$ is unknown for such cases, however, we
could not estimate $\eta$.

We note that spurious changes in LOS fields over a magnetogram FOV,
due, e.e., to artifacts arising from orbital motion in the case of HMI
(or any other cause) will introduce errors into the bias velocity
estimate: our approach attributes changes in unsigned LOS flux to
emergence / submergence, so changes unrelated to actual emergence /
submergence will introduce error.  Are the known artifacts in HMI
field strengths significant?  Probably not, since the orbital
variations occur on timescales much slower than the evolution over
$\Delta t =$ 24 min. that we analyze.  \citet{Liu2012b} report changes
in mean $|B_{\rm LOS}|$ of approx. 150 Mx cm$^{-2}$ peak to peak,
i.e., over 12 hr. (see their Figure 5).  This rate of change
corresponds to a change of 5 Mx cm$^{-2}$ over 24 min., which is at
the noise level of an individual LOS measurement. This implies these
slow variations in LOS field strength would be dominated by
fluctuations from noise, or faster evolution.

\subsection{Empirical Estimates of Uncertainties}
\label{subsec:uncertainties}

After filtering azimuths and removing Doppler velocities due to solar
rotation and convection, we interpolated the data to sub-pixel
accuracy for two purposes: (i) image co-registration, to remove
frame-to-frame jitter, primarily from quantized changes in the cutout
window boundaries (determined by pixel size) imposed on the smooth
rotation of the underlying solar features; and (ii) Mercator
projection of the spherical data onto a uniform grid in a Cartesian
frame.  While interpolation propagates errors from
pathological pixels, and can introduce other artifacts (e.g.,
smoothing) into the data, it should nonetheless be undertaken,
since significant whole-frame shifts were found (some frame-to-frame
shifts were on the order of 1 pixel, and cumulative shifts showed both
secular evolution and jumps) in the $x$ and $y$ directions by Fourier
cross-correlation in the $B_{\rm LOS}$ image sequence.

Since azimuth filtering and interpolation modify the input data, we
characterized uncertainties in the dataset empirically, instead of
relying upon error arrays produced in the HMI processing pipeline.  To
do so, we constructed histograms of the LOS and transverse fields,
$B_{\rm LOS}$ and $B_{\rm trs}$ respectively, at each time step, and
statistically characterized these distributions.\footnote{We created
  animations of these histograms for the data sequence, which are
  online at
  http://solarmuri.ssl.berkeley.edu/$\sim$welsch/public/data/AR11158/Noise/
}


In each distribution of $B_{\rm LOS}$, we interpret the weak-field
core of distribution as a measure of noise (e.g.,
\citealt{DeForest2007}), which we quantified by fitting the core of
the distribution ($\pm 6$ Mx cm$^{-2}$) with a Gaussian of width
$\sigma_{\rm L}$, and by finding the flux density at half max (HM) in
the left and right wings.  Statistically, the mean and standard
deviation of fitted $\sigma_{\rm L}$'s were $4.1 \pm 0.3$ Mx cm$^{-2}$
and combined left- and right-widths at HM were $4.7 \pm 0.5$ Mx
cm$^{-2}$, but fluctuations with a periodicity around $ 6$ hours were
observed.  In SDO's geosynchronous orbit, (i) Doppler motion toward
and away from the Sun and (ii) thermal cycling due to the Earth's
day-side albedo would both exhibit 12-hour periodicities, and would be
a quarter cycle out of phase, so could combine to produce variations
in instrumental response with such a characteristic 6-hour
periodicity.  This analysis roughly agrees with the HMI Team's
estimate of the uncertainty of $6$ Mx cm$^{-2}$, and suggests that the
uncertainty in the LOS field component assumed above, 20 Mx cm$^{-2}$,
probably overestimates the true uncertainty.

Since $B_{\rm trs}$ is a magnitude, it is necessarily positive, and we
find the distribution peaked away from zero.  If the distributions of
each of the three observed field components were Gaussian, with the
same width $\sigma$, then the distribution of transverse field
strengths would be that of a 3D Maxwell distribution, $\propto B_{\rm
  trs}^2 \exp(-B_{\rm trs}^2/2\sigma^2)$.  In fact, we find that a
Gaussian displaced from zero (fitted to data within $\pm 25$ Mx
cm$^{-2}$ of the tallest peak) consistently fit each distribution of
$B_{\rm trs}$ more closely (as quantified by $\chi^2$ over the whole
distribution) than a 3D (or 2D) Maxwellian.  It should be noted,
however, that the shape of the distribution varied substantially over
the observing sequence --- e.g., exhibiting two peaks at some time
steps --- so the Gaussian fits, although better than the Maxwellians,
were still poor in many cases.  The means and standard deviations of
the Gaussians' fitted centroids $B_{\rm t0}$ and widths $\sigma_{\rm
  t}$ were $81 \pm 17$ Mx cm$^{-2}$ and $29 \pm 6$ Mx cm$^{-2}$,
respectively.  Both also exhibited periodicities, though with more
complicated temporal structure than with fits to the distribution of
$B_{\rm LOS}$, with the clearest period at $\sim 24$ hr.  This
analysis suggests that our assumed uncertainty in the transverse field
component, 90 Mx cm$^{-2}$, is reasonable.

\section{Summary}
\label{sec:summ}

We have demonstrated a technique to estimate any spatially-constant
bias velocity $v_0$ present in measured photospheric Doppler
velocities using a sequence of observations of the vector magnetic
field near disk center.  Such biases can arise from instrumental
effects (e.g., thermal variations in instrument components) or from
the convective blueshift \citep{Dravins1981}.  Our technique uses
consistency between changes in LOS flux near PILs close to disk center
and the transport of transverse flux implied by Doppler velocities
along those PILs.

We also noted that an estimate of the bias velocity can be made more
simply, from the median of Doppler velocities along all pixels in LOS
PILs close to disk center.  This approach does not require measurement
of the transverse field; only an LOS magnetogram and a Dopplergram are
required.  One need only identify PILs, and take the median Doppler
velocity on all PIL pixels.  While this approach ignores our
flux-matching constraint in equation (\ref{eqn:ideal_equiv_fd}), and
could therefore, in principle, be biased by a strong episode of flux
emergence or submergence, we only found evidence of a very weak
bias.

Accurately calibrated Doppler velocities can be used to better
understand photospheric dynamics prior to and during CMEs
\citep{Schuck2010}, to improve estimates of the photospheric
electric field and thereby estimate the photospheric Poynting flux
\citep{Fisher2012b}, and to differentiate between possible mechanisms
of flux cancellation ($\Omega$-loop submergence, U-loop emergence, or
reconnective cancellation).  

For the HMI dataset we studied, we found clear evidence of bias
velocities, though estimates of $v_0$ ranged from less than 100 m
s$^{-1}$ to $\sim 500$ m s$^{-1}$ over the course of the run.  Only
the larger estimates of $v_0$ --- in the range of 350 $\pm$ 85 m
s$^{-1}$ --- were consistent with a majority of PILs being sites of
emergence, in accordance with the substantial increase in total
unsigned magnetic flux over the duration of our data set.  

Unfortunately, flux budgets based upon our estimates of the rates of
emergence and submergence are very noisy, and likely subject to large
errors.  Consequently, improving our approach to constraining the bias
velocity, or identifying new constraints, would be desirable.
We plan a careful investigation of systematic sources of uncertainty
in our method (e.g., better quantitative characterization of flux
cancellation and changes in LOS fluxes near PILs), or possible bias
(e.g., greater errors in shorter vs. longer PILs, or weak-field
vs. strong-field PILs).

It would also be worthwhile to compare results of our electromagnetic
calibration with calibrations based upon fitting the center-to-limb
component of full-disk HMI measurements of Doppler velocities
\citep{Snodgrass1984, Hathaway1992, Hathaway2000, Schuck2010}.
As noted above, center-to-limb fitting has a major physical
uncertainty: the physics of the center-to-limb variation in Doppler
shift in the particular spectral line used by HMI (or any other
spectral line) involves detailed interactions between height of
formation, the height of convective turnover, the variation with
viewing angle of the average convective flow speed, and the variation
with viewing angle of optical depth \citep{Carlsson2004}.  Diverging
flows tangent to the photosphere in granules can, depending upon
optical depth in granules at the formation height of the line, produce
a convective blueshift toward the limb (e.g., \citealt{Takeda2012}),
because diverging flows on the near sides of granules approach the
viewer, while receding flows on the far sides of the granules are at
least partially obscured by the optical depth of the granules (see
Figure \ref{fig:c2l}).  Hence, finding the center-to-limb shift in a
line does not imply that the line does not still possess an overall
additional shift.

We note that, in addition to the Dopplergram data analyzed here, the
HMI Team also produces independent estimates of Doppler velocities in
the Milne-Eddington inversion process.  While comparisons between ME
and Dopplergram velocities show them to be quite similar (Yang Liu,
private communication 2011), these independent measurements can
provide additional information, e.g., better characterization of
uncertainties.

In addition, as a step in our data reduction, we implemented a
procedure to correct a few episodes of single-frame fluctuations in
the direction of the transverse magnetic field vector, by nearly
180$^\circ$.  We attributed these flips-followed-by-reversals to
errors in resolution of the 180$^\circ$ ambiguity in the transverse
field.  As discussed in further detail above, our automated procedure
reverses suspicious transient flips by imposing temporal continuity:
among pixels with jumps in azimuth larger than 120$^\circ$ at frame
$i$, it reverses any azimuths that increase the average of the four
unsigned differences in azimuth between frame $i$ and the nearest $\pm
2$ frames.  This relatively simple algorithm corrected the episodes of
suspicious azimuth changes that we noticed.

\acknowledgements Data and images are courtesy of NASA/SDO and the HMI
science teams.  We thank Stanford University's Yang Liu, Jesper Schou,
and Sebastien Couvidat, and NWRA's K.D. Leka for readily answering
queries about HMI data processing.  SDO/HMI is a joint effort of many
teams and individuals, whose efforts are greatly appreciated.  This
research was funded by the NASA Heliophysics Theory Program (grant
NNX11AJ65G), the NASA Living-With-a-Star TR\&T Program (grant
NNX11AQ56G), and support from NSF via grant AGS-1048318 and the
National Space Weather Program (grant AGS-1024862).  X. Sun is
supported by NASA Contract NAS5-02139 (HMI) to Stanford University.
The authors are grateful to US taxpayers for providing the funds
necessary to perform this work.

\appendix

\section{Geometric Effects Away from Disk Center}
\label{app:offcm}

\begin{figure}[!htb] 
  \centerline{\psfig{figure=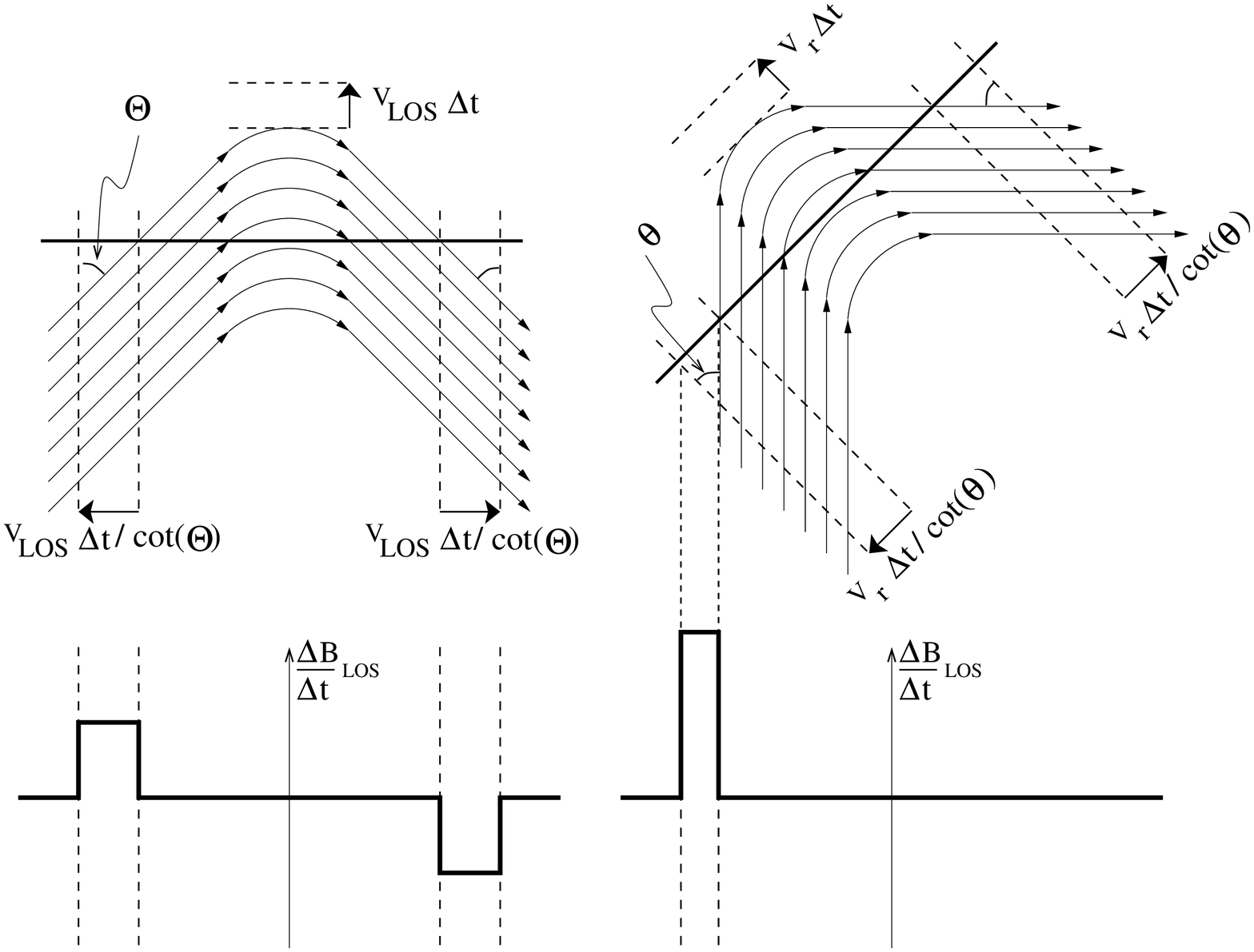,width=7.0in}}
\caption{\footnotesize \textsl{Upper left and right panels depict a
    highly idealized emerging flux configuration, assumed to be
    invariant in the direction into the page, viewed in two different
    orientations with respect to an observer, who is assumed to be
    looking down from above the top of this figure.  In the upper left
    panel, measurements of Doppler velocities on the radial-field PIL
    and transverse fields at the PIL can be used to infer the rate of
    flux emergence.  In the upper right panel, the magnetic field at
    the radial-field PIL has a component along the LOS, so flows
    parallel to $\bvec$ (which do not drive flux emergence) can
    contribute to Doppler shifts, implying the Doppler velocity cannot
    unambiguously determine the rate of flux emergence at this PIL.
    In both cases, flux emergence can be detected from increases in
    total unsigned LOS flux (the curves plotted in the bottom left and
    right panels), and quantitatively estimated from changes in the
    total unsigned radial magnetic flux.}}
\label{fig:offcm}
\end{figure}

We have asserted that the calibration method we describe here is only
valid near disk center.  To better understand why, we consider a
particular configuration of emerging magnetic flux in Figure
\ref{fig:offcm}, where it is illustrated in two different orientations
with respect to the viewer.  This configuration is highly idealized:
we assume the configuration is invariant in the direction into the
page, and that emerging field lines have the same shape as field lines
that have already emerged.  In both panels, the viewer is assumed to
look down from above the top of the figure.  

In the upper left panel, a plane tangent to the photosphere at the
radial-field PIL is normal to the LOS.  At this particular viewing
angle, the photospheric normal direction and the LOS coincide.  Away
from the PIL, the field is uniformly tilted from the LOS by the
inclination angle $\Theta = 45^\circ.$ In the upper right panel, the
normal to the plane tangent to the photosphere at the radial-field PIL
is tilted with respect to the LOS by an angle $\theta = 45^\circ$,
which is also the tilt of the legs of the emerging flux system with
respect to the photosphere.  
%

We assume the flux emergence is ideal, such that the electric field
obeys equation (\ref{eqn:ideal}), and flux is therefore frozen to the
plasma.  The rate of flux emergence by a radial flow can therefore be
inferred from observations of the upper-left panel in two ways: (1)
from the increase in the the total unsigned LOS flux (the integrated
absolute value of the curve plotted in the lower left panel); and (2)
from the product of the transverse field at the PIL and the Doppler
signal at the PIL.  The equivalence of these measures of the rate
ideal flux emergence underlies the Doppler calibration method
presented in this paper: we assume any inconsistency between (1) and
(2) is due to a bias in the Doppler zero point.

In contrast, flux emergence can only be unambiguously inferred by
observations of the upper right panel one way: from the increase in
the the total unsigned flux (either LOS flux or radial flux).  This is
because the magnetic field at the radial-field PIL has a component
along the LOS, implying Doppler shifts at this PIL might contain a
contribution from flows parallel to $\bvec$ there.  Flows parallel to
$\bvec$, however, do not drive flux emergence, and cannot be
distinguished from Doppler shifts due to flows perpendicular to $\bvec$.

It is also clear that the relative orientation of the
negative-polarity flux and the LOS implies that an observer would not
see an increase in negative-polarity LOS flux as emergence proceeds.
Hence, observations of the emergence of tilted flux in LOS
magnetograms can exhibit strong flux imbalance.  Measurements of
changes in the radial flux, from vector magnetograms, would be
balanced in this idealized case.

%



\end{document}